\documentclass[]{article}
\usepackage{style}

\usepackage[utf8]{inputenc} 
\usepackage[T1]{fontenc}    
\usepackage{hyperref}       
\usepackage{url}            
\usepackage{booktabs}       
\usepackage{amsfonts}       
\usepackage{nicefrac}       
\usepackage{microtype}      
\usepackage{lipsum}
\usepackage{fancyhdr}       
\usepackage{graphicx}       
\usepackage{multicol}
\usepackage{longtable}
\usepackage{tabularx}
\usepackage{float}
\usepackage{caption}
\usepackage{setspace} 
\usepackage{placeins}
\usepackage[
backend=biber,
style=ieee,
citestyle=numeric-comp
]{biblatex}

\usepackage[utf8]{inputenc}
\usepackage[T1]{fontenc}

\DeclareUnicodeCharacter{1EC7}{e}

\AtEveryBibitem{\clearfield{address}}
\AtEveryBibitem{\clearlist{address}}

\AtEveryBibitem{\clearfield{place}}
\AtEveryBibitem{\clearlist{place}}

\AtEveryBibitem{\clearfield{issn}}
\AtEveryBibitem{\clearlist{issn}}

\AtEveryBibitem{\clearfield{series}}
\AtEveryBibitem{\clearlist{series}}

\AtEveryBibitem{\clearfield{month}}
\AtEveryBibitem{\clearlist{month}}

\AtEveryBibitem{\clearfield{isbn}}
\AtEveryBibitem{\clearlist{isbn}}

\AtEveryBibitem{\clearfield{language}}
\AtEveryBibitem{\clearlist{language}}

\AtEveryBibitem{\clearfield{doi}}
\AtEveryBibitem{\clearlist{doi}}

\AtEveryBibitem{%
  \ifentrytype{misc}
    {%
    }
    {%
      \clearfield{url}
      \clearlist{url}
    }%
}

\AtEveryBibitem{\clearfield{extra}}
\AtEveryBibitem{\clearlist{extra}}

\AtEveryBibitem{\clearfield{note}}
\AtEveryBibitem{\clearlist{note}}

\AtEveryBibitem{%
  \ifentrytype{online}
    {}
    {\clearfield{urlyear}\clearfield{urlmonth}\clearfield{urlday}}}

\graphicspath{{media/}} 

\title{Social Media Use is Predictable from App Sequences:
 Using LSTM and Transformer Neural Networks to Model Habitual Behavior
}

\author{
Heinrich Peters            \\
Columbia University\\
\texttt{\ hp2500@columbia.edu}\\
\And
Joseph B. Bayer\\
Ohio State University \\
\texttt{\ bayer.66@osu.edu}\\
\And
Sandra C. Matz\\
Columbia University\\
\texttt{\ sm4409@columbia.edu}\\
\AND
Yikun Chi\\
Stanford University\\
\texttt{\ yikunchi@stanford.edu}\\
\And
Sumer S. Vaid\\
Harvard University\\
\texttt{\ svaid@hbs.edu}\\
\And
Gabriella M. Harari\\
Stanford University\\
\texttt{\ gharari@stanford.edu}\\
}

\begin{document}

\maketitle


\vspace{3cm}

\begin{abstract}
The present paper introduces a novel approach to studying social media habits through predictive modeling of sequential smartphone user behaviors. While much of the literature on media and technology habits has relied on self-report questionnaires and simple behavioral frequency measures, we examine an important yet understudied aspect of media and technology habits: their embeddedness in repetitive behavioral sequences. Leveraging Long Short-Term Memory (LSTM) and transformer neural networks, we show that (i) social media use is predictable at the within and between-person level and that (ii) there are robust individual differences in the predictability of social media use. We examine the performance of several modeling approaches, including (i) global models trained on the pooled data from all participants, (ii) idiographic person-specific models, and (iii) global models fine-tuned on person-specific data. Neither person-specific modeling nor fine-tuning on person-specific data substantially outperformed the global models, indicating that the global models were able to represent a variety of idiosyncratic behavioral patterns. Additionally, our analyses reveal that the person-level predictability of social media use is not substantially related to the frequency of smartphone use in general or the frequency of social media use, indicating that our approach captures an aspect of habits that is distinct from behavioral frequency. Implications for habit modeling and theoretical development are discussed.
\end{abstract}
\vspace{1cm}

\keywords{Social Media, Habits, User Modeling, LSTM, Transformer, Neural Networks}




\newpage
\section{Introduction}

Mobile app use is ubiquitous in our daily lives. On average, US smartphone users spend 4.5 hours per day on their phones \cite{statista_us_2024}, maintaining relationships, conducting business, and seeking information or entertainment. Among the long list of technology-mediated behaviors, social media use deserves special attention, not only because social media platforms are used by billions of people every day but also because they have been shown to impact almost every aspect of our lives. For example, social media use has revolutionized social interaction \cite{coyle_social_2008, diehl_political_2016, licoppe_are_2005, zeitzoff_how_2017}, changed the consumption of information and news \cite{gil_de_zuniga_effects_2017, lee_effects_2017, vermeer_online_2020, walker_news_2021}, and been linked to a wide range of indicators of well-being \cite{allcott_welfare_2020, bayer_building_2022, brailovskaia_less_2020, fox_dark_2015, hunt_no_2018, primack_social_2017, verduyn_passive_2015, vaid_variation_2024}.

Here, we investigate the predictability of social media use through the theoretical lens of media and technology habits. Habits are commonly defined as repeated behaviors that are automatically initiated by contextual cues, such as locations, situational elements, and preceding actions or activities \cite{anderson_habits_2021, verplanken_technology_2018, gardner_habit_2019, schnauber-stockmann_process_2019, tokunaga_media_2020, verplanken_beyond_2006, wood_psychology_2016}. Thus, habitual behaviors are performed in a semi-conscious manner in which individuals follow a rehearsed script once triggered by the contextual cue \cite{bayer_consciousness_2016}. Habits are formed as individuals learn implicit associations between cues and rewarded responses, which become automatic and independent of rewards over time \cite{bayer_building_2022, wood_habit_2017, orbell_automatic_2010, gardner_habit_2012}. Social media users, for instance, may develop habits by associating specific cues (e.g., notifications, boredom, or preceding app use) with subsequent user behaviors (e.g., reaching for one's phone or clicking on an app or notification) in particular contexts. These cues then trigger relevant behaviors automatically, even if the initial rewards are removed, making behavior more efficient by conserving mental energy \cite{bayer_building_2022, wood_psychology_2016}. The concept of habits provides a promising perspective on media and technology use because it accounts for reward-driven as well as automatic repetitive behavior and is grounded in a rich theoretical tradition including behavioral \cite{naab_habitual_2016, nebe_characterizing_2024}, cognitive \cite{wood_new_2007, wood_psychology_2016}, and neuropsychological \cite{mendelsohn_creatures_2019, amaya_neurobiology_2018} approaches, as well as links to the literature on well-being \cite{docherty_digital_2021, docherty_facebooks_2020, larose_social_2011}.

Past research indicates that social media use is, in large part, habitual behavior \cite{anderson_habits_2021, bayer_building_2022}. For instance, one might habitually check one's phone while waiting for a bus on the way to work, respond to messages on Facebook before going to bed, or check Instagram by default when unlocking one's phone. Consequently, sequences of people’s interactions with their devices tend to exhibit regularities over time. For example, social interaction and content consumption \cite{peters_context-aware_2023}, instant messaging \cite{peters_predicting_2022}, and in-app action sequences \cite{liu_characterizing_2019}, follow predictable patterns that have been characterized as habitual in nature. Moreover, there is a growing body of predictive work in human-computer interaction and user modeling, suggesting that smartphone user behaviors, including app engagement, are predictable to varying degrees from factors like device states, temporal context, location, and preceding actions \cite{baeza-yates_predicting_2015, huang_predicting_2012, natarajan_which_2013, parate_practical_2013, xia_deepapp_2020, xu_predicting_2020}.

While these findings demonstrate regularities in smartphone user behaviors that can be interpreted as habitual, the behavioral science literature has taken a different approach, often relying on relatively simple behavioral measures (e.g., behavioral frequency, reward devaluation \cite{nebe_characterizing_2024, mendelsohn_creatures_2019}) or self-report measures to assess habits (e.g., Self-Report Habit Index \cite{verplanken_reflections_2003}, Self-Report Behavioral Automaticity Index \cite{gardner_towards_2012}, Unified Theory of Acceptance and Use of Technology questionnaire \cite{venkatesh_consumer_2012}, Response Frequency Measure of Media Habit \cite{naab_habitual_2016}). 
These methods have proven useful in past research, but they are also limited in at least two important ways. First, people tend to be inaccurate in reporting past behavior, especially social media use \cite{boyle_systematic_2022}. Notably, such recall biases might be especially pronounced in the case of mobile and social media habits, as they often occur automatically and without conscious engagement in the midst of daily life \cite{bayer_connection_2016, bayer_consciousness_2016, bayer_texting_2012}. Second, questionnaire-based methods are ill-equipped to detect more granular behavioral patterns, such as sequences of device interactions. Significantly, the same limitation applies to simple behavioral measures, which ignore the embeddedness of behaviors in preceding and subsequent activities along with surrounding contextual factors. Integrating more objective, granular, and context-sensitive methods grounded in digital trace data could provide a more holistic and accurate representation of media habits.

Here, we shift focus from self-report measures to tracking behavioral sequences via smartphone logs. Specifically, we build on recent research positing that repetitive behavioral sequences constitute an important, yet understudied, aspect of media and technology habits \cite{roffarello_understanding_2021}. Under this scope, we can capture media habits as the extent to which a focal behavior (e.g. the use of a particular social media app) can be predicted from the history of preceding behaviors (e.g., the use of other applications). Hence, recurring sequences can indicate the extent to which an app (or combination of apps) acts as a contextual trigger for other app selections, aligning with past habit research on the role of preceding actions as contextual cues \cite[see][]{wood_habit_2017}. This operationalization of media habits offers several important advantages compared to traditional approaches to measuring habit processes. First, we are able to directly extract habits from objectively trackable smartphone user behaviors in the real world, thereby circumventing the previously outlined response biases and artificial scenarios. Second, the reliance on sequences of behavior allows us to paint a more nuanced picture of habitual behaviors than merely focusing on behavioral frequencies.

The modeling of person-specific behavioral processes has a long tradition in the behavioral sciences \cite[e.g.,][]{hamaker_idiographic_2009, haslbeck_recovering_2022, lally_how_2010, molenaar_manifesto_2004, molenaar_new_2009, ram_binding_2023}, but much of this work has relied on autoregressive models \cite{haslbeck_recovering_2022} or classic machine learning techniques \cite{beck_personalized_2022, buyalskaya_what_2023}. While these approaches can offer insights into the structure of behavioral sequences, the increasing adoption of deep learning approaches that are specifically designed to represent complex time series, such as Long-Short-Term-Memory (LSTM) \cite{hochreiter_long_1997, schmidhuber_deep_2015} or transformer neural networks \cite{vaswani_attention_2017}, opens the door for more sophisticated modeling approaches \cite{ram_binding_2023}.

Given the fact that habits are - by definition - repetitive, the present paper examines the predictability of social media habits using LSTM and transformer neural network models. In doing so, it aims to answer the following research question: Does social media use follow predictable patterns that can be characterized as habitual? For this purpose, we first examine the extent to which social media use is predictable from intensive longitudinal app log data at the between and within-person levels and the extent to which the predictability of social media use varies across individuals. Second, we explore how the predictability of social media use relates to behavioral frequency measures and other properties of the underlying training data. Third, we examine the size of the relevant behavioral context window by analyzing the relationship between predictive performance and the length of the behavioral sequences available to the models. Our analyses are based on the predictive performance of (i) global models trained on pooled data from all users, (ii) person-specific models trained from scratch on person-level data, and (iii) global models fine-tuned on person-level data. The present study thus employs sequential machine learning approaches to explore a new operationalization of habits that is rooted in the predictability of objectively trackable behavior.

\section{Method}
\subsection{Participants and Procedure}
We collected data from a sample of N=182 adults (final N=99 after data cleaning; see below) between mid-January and early March 2021. Participants were recruited through Prolific, and all were located in the US. All participants were Android phone users. The current study focuses on Android app logs (time-stamped records of interactions with various smartphone apps), collected through the Screenomics app \cite{reeves_screenomics_2021}. The app logs were recorded through event-based sampling each time an app was used. Each app was identified by its associated Android package name, a unique identifier used by the Android operating system and Google Play Store. All participants gave informed consent. The study received approval under the Stanford University IRB (Protocol \#48234).

\subsection{Data Preprocessing}
In order to standardize the observation period per user, we discarded users with less than 14 full days of participation, and we truncated the participation period to 14 days for those who had participated for more than 14 days. From the app log sequences, we discarded instances of the Screenomics app, which was used for data collection (0.56\% of observations), as well as empty events for which no app was logged (1.65\% of observations). We also discarded logs associated with navigational (e.g., Launcher) and system UI processes (37.06\% of observations), as these events represent engagement with the Android operating system rather than direct usage of apps. Finally, in order to guarantee sufficient training data at the person level, we removed users with less than 1,000 logged app sessions over the course of the observation period and users with minimal social media use (<5\% of observations). Through this process, we reduced the participant sample size from N=182 to N=99, and the number of observations from 585,576 to 282,639.

The target variable, social media use, was derived directly from the app logs using a list of social media apps (e.g., Facebook, Instagram, TikTok, Snapchat; the full list can be found in SI A), based on the conceptualization of social media proposed by Bayer et al. \cite{bayer_social_2020}. Messaging apps like WhatsApp, Telegram, and Messenger were not considered social media apps for the purpose of the study. 

\subsection{Modeling}
We utilized two neural network architectures that are specifically designed to capture patterns in sequential data: LSTM \cite{hochreiter_long_1997, schmidhuber_deep_2015} and transformer neural networks \cite{vaswani_attention_2017}. LSTM neural networks are a type of recurrent neural network designed to represent long-term dependencies in sequential data, making them particularly effective for tasks like language processing or time series analysis \cite{graves_hybrid_2013, han_ese_2017,siami-namini_performance_2019}. Transformer neural networks are a type of architecture that relies on self-attention mechanisms \cite{vaswani_attention_2017}, enabling the processing of entire sequences of data simultaneously, making them highly efficient and effective for tasks like natural language understanding and generation \cite{devlin_bert_2019, openai_gpt-4_2023}. In addition, they also excel at learning from other types of sequential data, such as health records \cite{li_behrt_2020} or life events \cite{savcisens_using_2024}. 

In order to standardize the prediction task across individuals, we framed it as a binary classification task, predicting whether the next app opened by the user would be a social media app or not. For the main analyses, we constructed a sequence of the 20 preceding actions for each logged app use event, irrespective of the temporal distance between events or whether they were separated by a period of inactivity. This was done separately within each of the training, validation, and testing sets in order to prevent information leakage across the splits. For the small subset of actions that did not have a sufficient number of preceding data points, we pre-padded the sequences with zeros. The app logs were fed into an embedding layer, which – in turn – fed into one or more LSTM or transformer layers and a single fully connected layer before the output layer. The output layer was a single neuron with a sigmoid activation function.

We first trained general, global models on the combined data of all participants. Second, following an idiographic modeling approach, we trained person-specific models for each individual in the dataset. Third, we implemented a hybrid strategy by fine-tuning the global model with person-specific data for each participant. With the exception of the fine-tuning stage, we tuned the neural network architecture and regularization parameters for each model in order to avoid spurious differences in model performance as a result of a mismatch between model hyperparameter settings and idiosyncratic properties of each person’s data. The tuned hyperparameters include architectural hyperparameters such as the dimensionality of the embedding layer ([5, 50]), the number ([1, 3]) and dimensionalities ([4, 64]) of the LSTM, transformer layers, and dense layers. In order to combat overfitting, we also tuned various regularization parameters such as dropout rate ([0.2, 0.5]) and recurrent dropout rate ([0.2, 0.5]), as well as L1 ([1e-5, 1e-3]) and L2 ([1e-4, 1e-2]) norms tuned separately for recurrent, transformer, and dense layers. Finally, we tuned the learning rate ([1e-4, 1e-2]) that was applied to the Adam optimizer \cite{kingma_adam_2017}. Hyperparameter search was performed using Bayesian Optimization \cite{falkner_bohb_2018} with 20 iterations and five random starting points. The optimization target was the area under the receiver operating characteristic curve (AUC). The AUC was used as the optimization target because it is insensitive to changes in the class distribution. This characteristic makes it especially useful in imbalanced datasets where one class outnumbers the other. A detailed description of the hyperparameter spaces can be found in SI B. 

A temporal train-validation-test split was performed for each individual, such that the first section of the data was allocated to the training set (50\% of observations), the second section was allocated to the validation set (25\% of observations), and the final section was allocated to the testing set (25\% of observations). The reason for choosing a temporal split was to test conservatively whether a model trained on data from the past would be predictive in the future. The training set was used to fit the model, the validation set was used for hyperparameter tuning, and the testing set was used to evaluate model performance.  For the global model, the splits were comprised of the union of person-level splits across participants. In other words, we maintained the same splits that were used at the person level and combined all individual training/validation/test sets to create overall training/validation/test sets. Each model was trained for up to 1,000 epochs with a batch size of 1,024 and early stopping with a patience parameter value of 5 epochs. We chose the relatively low patience value to avoid overfitting. The stopping criterion was the binary cross-entropy loss computed on the validation set. In order to account for class imbalances in the testing set, we employed the AUC as the main evaluation metric. An AUC value ranges from 0 to 1, where a value of 0.5 indicates a model with no discriminative power (equivalent to random guessing or a majority class classifier), and a value of 1 signifies a perfect model with complete separability of classes. The AUC metric is advantageous because it is independent of the decision threshold and provides a single number summary of the model's performance across all possible probability thresholds, making it a robust measure for comparing different classifiers. Other commonly used binary classification metrics can be found in SI C-H. 

Aside from the neural network models described above, we utilized two classic ML algorithms as additional benchmarks. Specifically, we trained regularized logistic regression \cite{tibshirani_regression_1996} as well as random forest \cite{breiman_random_2001} models, neither of which performed on par with the neural network models.

All analyses were conducted using the Tensorflow Python library \cite{abadi_tensorflow_2016}, the Kerastuner Python library \cite{omalley_keras_2019}, and the Scikit-Learn \cite{pedregosa_scikit-learn_2018} library. The code used to generate the results is available on this paper’s OSF page (\href{https://osf.io/rkswe/}{https://osf.io/rkswe/}).

\section{Results}
\subsection{Descriptive Statistics}
\begin{figure}[t]
  \centering
  \includegraphics[width=1\textwidth]{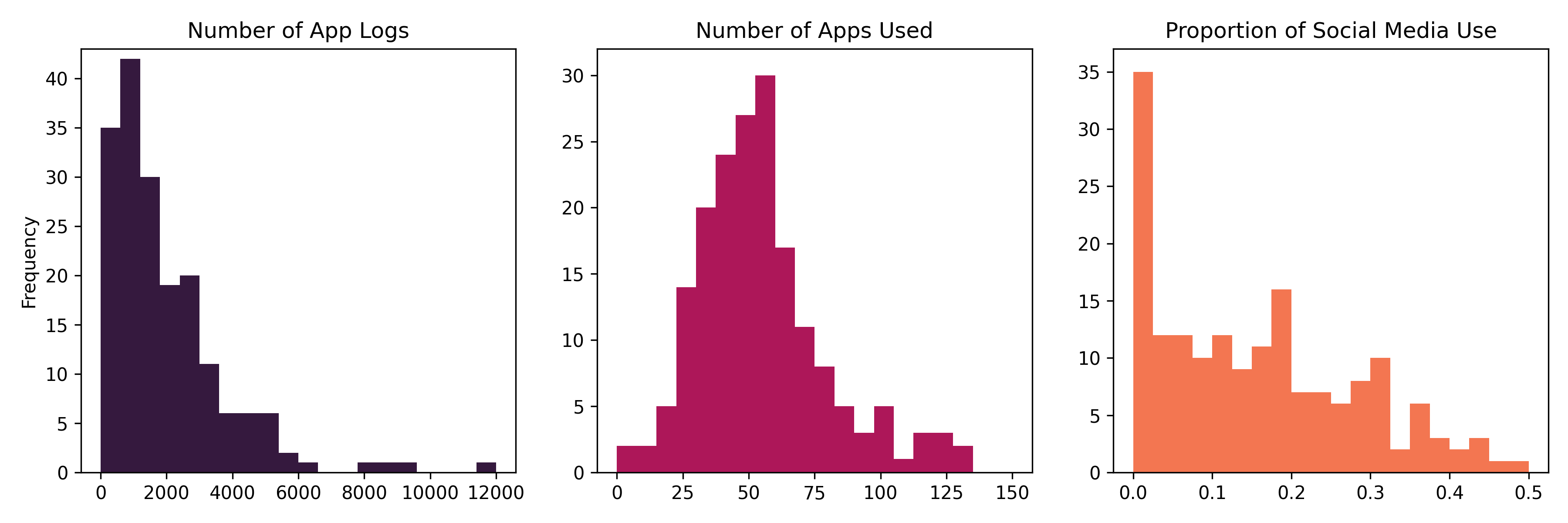}
  \includegraphics[width=1\textwidth]{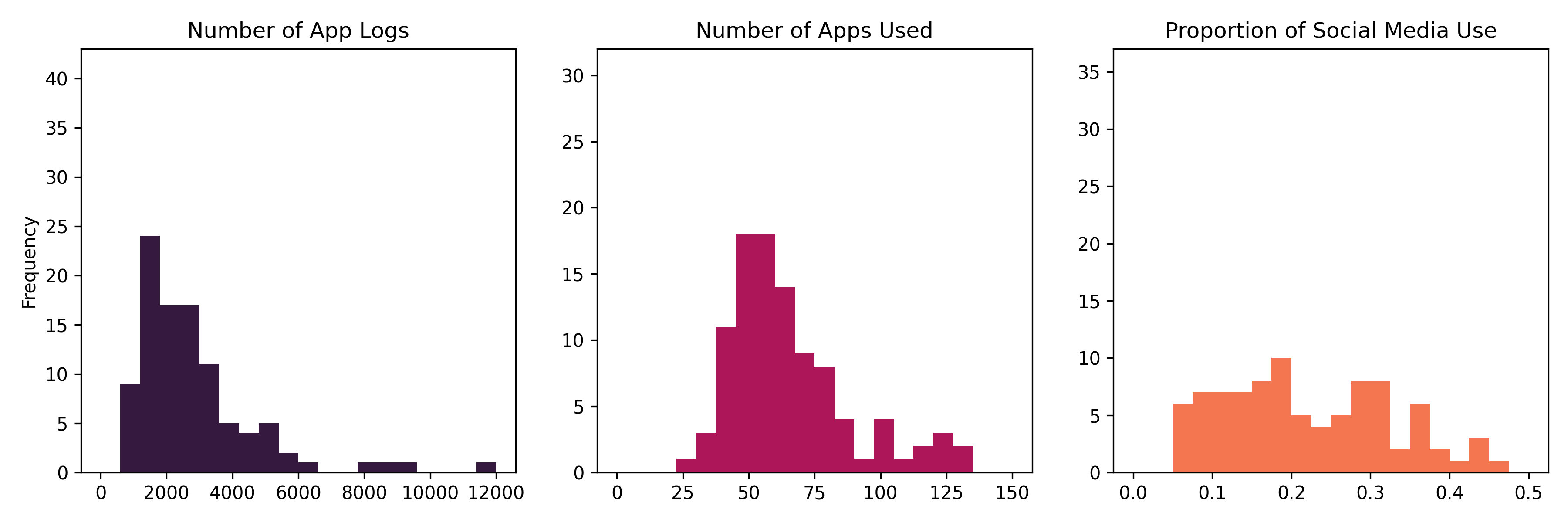}

  \caption[Frequency distributions of app use behaviors]
  {Frequency distributions across participants for the number of app logs, the number of distinct apps used during the observation period, and the proportion of social media sessions relative to other apps prior to excluding participants from the dataset pre (top) and post-cleaning (bottom).}
  \label{fig:habit_descriptives}
\end{figure}

On average, before excluding individuals from the analysis given cleaning criteria, each participant had 1980 logged app sessions (Median=1456.5, SD=1804.2). The average number of distinct apps (Android packages) used per person was 54.4 (Median=50.5, SD=24.5). The average proportion of app sessions that were social media sessions was 17.68\% (Median=15.31, SD=15.29). After cleaning, the average number of app sessions was 2856 (Median=2374, SD=1875.5), the average number of distinct apps (Android packages) used per person was 64.8 (Median=59, SD=23.1), and the average proportion of social media sessions was 25.33\% (Median=22.09, SD=14.43). A visual representation of the distributions can be found in Figure \ref{fig:habit_descriptives}.

The most commonly used apps were Google Chrome (10.07\% of observations, 9.34\% after cleaning), Facebook (5.04\% of observations, 6.17\% after cleaning), Google Mail (4.71\% of observations, 4.58\% after cleaning), Saumsung’s messaging app (4.64\% of observations, 4.63\% after cleaning), Google Quick Search Box (3.76\% of observations, 3.65\% after cleaning), Messenger (3.51\% of observations, 3.97\% after cleaning), Google’s messaging app (3.47\%, 2.91 after cleaning), Snapchat (3.13\% of observations, 3.88\% after cleaning), Instagram (3.07\% of observations, 3.77\% after cleaning), and Reddit (2.82\% of observations, 3.33\% after cleaning). A detailed list can be found in Table 1. In addition to individual app sessions, we also compiled a list of the most frequent app session 3-grams (unique sequences spanning 3 sessions each) along with their social media transition probabilities (the empirical probability of subsequent social media use). This information is available for the overall sample as well as examples of individual users in SI I. 

\begin{table}[ht]
   
\scriptsize

\noindent\begin{tabularx}{\textwidth}{Xrrrr}
\toprule
App & Proportion of Sessions & Proportion of Users & Proportion of Sessions & Proportion of Users \\
 & Pre-Cleaning &  Pre-Cleaning & Post-Cleaning & Post-Cleaning \\

\midrule
Google Chrome & 10.07 & 92.86 & 9.34 & 95.96 \\
Facebook & 5.04 & 52.20 & 6.17 & 68.69 \\
Google Mail & 4.71 & 84.62 & 4.58 & 88.89 \\
Samsung Messaging & 4.64 & 39.56 & 4.63 & 44.44 \\
Google Search & 3.76 & 87.36 & 3.65 & 91.92 \\
Messenger & 3.51 & 53.30 & 3.97 & 66.67 \\
Google Messaging & 3.47 & 39.56 & 2.91 & 36.36 \\
Snapchat & 3.13 & 34.62 & 3.88 & 49.49 \\
Instagram & 3.07 & 47.80 & 3.77 & 70.71 \\
Reddit & 2.82 & 40.11 & 3.33 & 54.55 \\
Discord & 2.34 & 28.02 & 2.81 & 35.35 \\
Twitter & 1.89 & 31.87 & 2.32 & 44.44 \\
Samsung Browser & 1.82 & 18.13 & 1.80 & 25.25 \\
Youtube & 1.49 & 73.08 & 1.58 & 85.86 \\
TikTok & 1.36 & 25.27 & 1.62 & 34.34 \\
Settings & 1.17 & 95.05 & 0.92 & 96.97 \\
Samsung Call UI & 0.98 & 47.80 & 1.00 & 56.57 \\
Gallery & 0.96 & 47.25 & 1.03 & 57.58 \\
Google Play Store & 0.96 & 92.31 & 0.81 & 96.97 \\
Spotify & 0.69 & 31.87 & 0.83 & 42.42 \\

\bottomrule
\end{tabularx}
\vspace{0.2cm}
    \caption[List of most frequently used apps]{List of most used apps by proportion of sessions and proportion of users who used the app at least once during the observation period in percent.}
    \label{tab:my_label}
\end{table}

\subsection{Predictability of Social Media Use}
\paragraph{Global Pre-Trained Models.}
We first trained a global model on the combined training sets of all individuals and performed a hyperparameter search using the combined validation sets of all individuals. Both the LSTM and the transformer model performed well on the combined test sets  ($AUC_{lstm}$=0.782, $AUC_{trans}$=0.773), suggesting that social media app use is moderately to highly predictable across individuals (for additional evaluation metrics, please refer to SI C).

\begin{figure}
  \centering
  \includegraphics[width=1\textwidth]{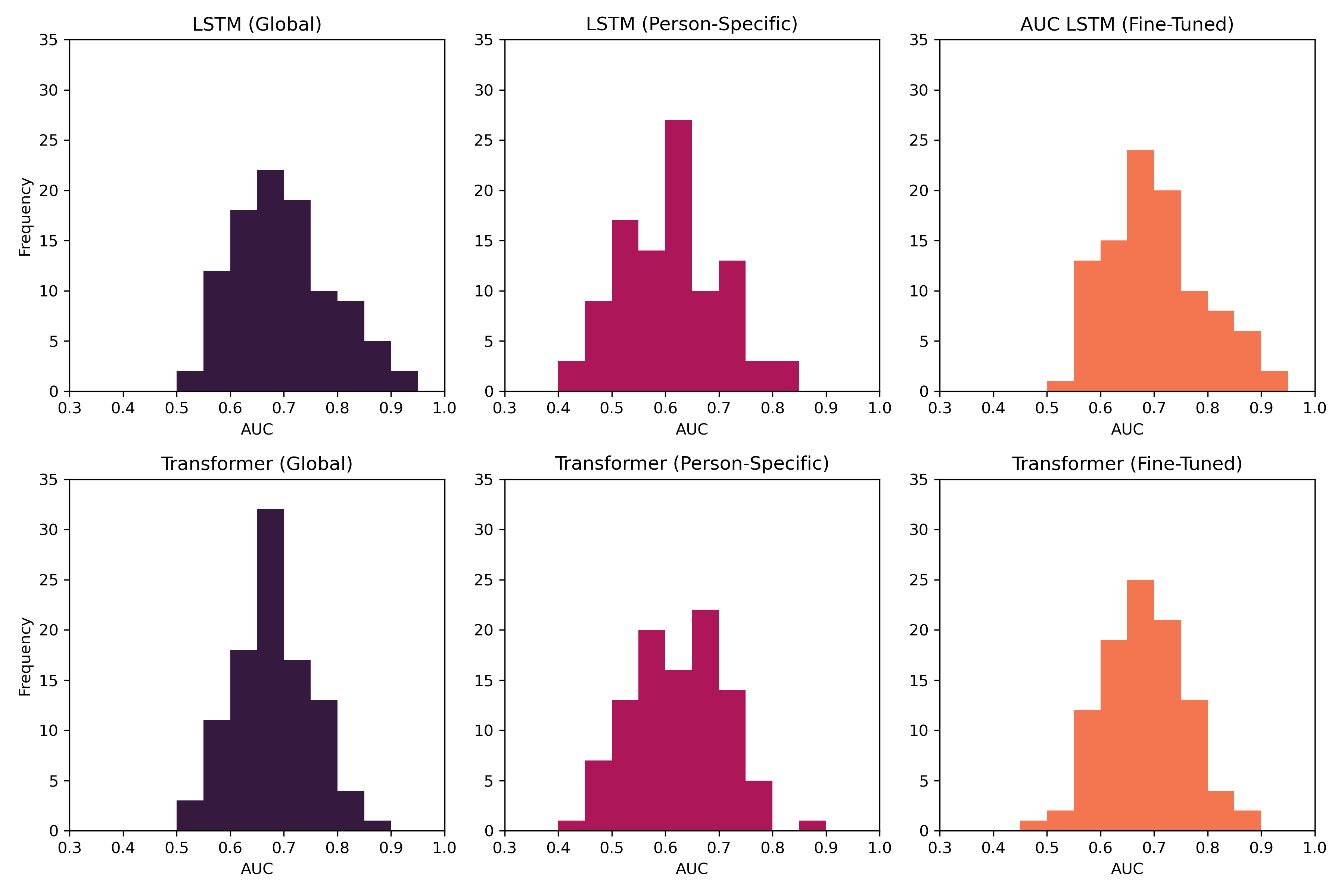}
  \caption[Distributions of model performance scores across participants]
  {Distributions of model performance scores (AUC) across participants for global (left), person-specific (middle), and fine-tuned (right) models.}
  \label{fig:auc_distributions}
\end{figure}

We then evaluated the global model on the individual test set of each participant to analyze the distribution of model performance scores across individuals. For this purpose, we computed separate AUC scores for each individual test set based on the predicted probability scores of the global models. We found considerable variance in the predictability of social media use across individuals for the global model. On average, the LSTM models yielded an AUC of 0.699 (SD=0.088, Min=0.515, Max=0.908), and the transformer models yielded an AUC of 0.679 (SD=0.072, Min=0.507, Max=0.853). The difference between global and person-level AUC scores indicates that the global model picks up on between-person variation in addition to within-person variation. A graphical representation of the distributions can be found in Figure \ref{fig:auc_distributions}.

To determine the sensitivity of the results to data cleaning decisions, we also evaluated an additional set of models trained on a less conservatively cleaned data set (i.e., including launcher and other system UI processes). While the predictive performance in these alternative models was generally higher than in the ones presented here (navigational events can be highly predictive of subsequent app use), the overall pattern of results was almost identical (see SI D). Additionally, we evaluated a set of models accounting for class imbalances in training by weighting the loss for each class with a value inverse to its frequency. The class weights were computed for each individual model based on the relative frequencies observed in its training set. While this approach led to slightly lower predictive performance in the case of transformer models, the overall pattern of results was almost identical to those of the main analysis as well (see SI E). Models trained on sequences only consisting of app sessions from the same day as the target session produced identical results, too (see SI F). Regularized logistic regression and random forest models, on the other hand, were outperformed by the neural network models (see SI G).


\paragraph{Person-Specific Modeling.}
Next, we trained a series of models for each participant to test whether idiosyncratic patterns of social media use would be better represented in person-specific models. Hyperparameter search was performed on person-level validation data, and model performance was evaluated on person-level test data. The splits were held constant with respect to those used for the global model. Similar to the global model, the person-specific models showed varying predictive performance across individuals. On average, the LSTM models yielded an AUC of 0.609 (SD=0.092, Min=0.414, Max=0.817), and the transformer models yielded an AUC of 0.627 (SD=0.087, Min=0.447, Max=0.861). The models did not outperform the global model on average, but for a small subset of participants (LSTM: 13.13\%  of participants; transformer: 26.26\% of participants), the individual models showed improved model performance. A graphical representation of the distributions can be found in Figure \ref{fig:auc_distributions}.

To test whether the relative strength of the global models compared to person-specific models was based on their ability to represent diverse behavioral patterns or simply because individuals exhibited very similar behavioral patterns in the first place, we examined the generalizability of person-specific models across participants. If the person-specific models were to generalize well across participants, this would indicate that their patterns of social media use were indeed similar. If the person-specific models did not generalize across individuals, we would conclude the inverse. This would suggest that the global models, too, were able to represent a multitude of idiosyncratic behavioral patterns.

We evaluated each person-specific model on the testing sets of all other individuals across the whole sample and then compared the average performance to the original results. We found that person-specific models did not generalize well across individuals with an average performance of AUC=0.526 for LSTM and AUC=0.538 for transformer models, both of which are considerably lower than their performance on within-person testing data and marginally beat the chance baseline of AUC=0.5. These results indicate that the person-specific models do not pick up on general behavioral patterns that are common across individuals. Instead, the strength of the models, including the pre-trained global models, seems to be driven by their ability to capture a multitude of individual behavioral patterns.

\paragraph{Fine-Tuning Global Models on Person-Specific Data.}
We fine-tuned the pre-trained global model to test whether personalized models could benefit from general representations of patterns of social media use and vice versa. For this purpose, we continued to train the pre-trained global models for up to 1000 additional epochs on the training data of each individual and performed early stopping using the binary cross-entropy loss on the individual’s validation set as the stopping criterion. In order to avoid overriding previously learned relationships, a low learning rate of 0.0001 was applied. This procedure resulted in a fine-tuned LSTM and a fine-tuned transformer model for each individual. On average, the fine-tuned LSTM models yielded an AUC of 0.702 (SD=0.089, Min=0.522, Max=0.908), and the fine-tuned transformer models yielded an AUC of 0.683 (SD=0.077, Min=0.497, Max=0.858). While the fine-tuned models outperformed the global model in most cases (LSTM: 68.68\%  of participants; transformer: 63.63\%  of participants), the fine-tuning only resulted in a very minor improvement over and above the global pre-trained model. A graphical representation of the distributions can be found in Figure \ref{fig:auc_distributions}.

In order to incorporate hyperparameter tuning into the fine-tuning approach, an alternative set of fine-tuned models was trained by freezing the model weights of the embedding layers, LSTM layers, and transformer layers while retraining the dense layers and output layers from scratch with another round of hyperparameter search (i.e., adjusting layer dimensions, regularization, and learning rate). This approach places a higher emphasis on finding appropriate hyperparameters at the cost of discarding the model weights of layers for which hyperparameters are tuned. The procedure did not result in improved model performance (see SI H). Taken together, the results show that the pre-trained global models already captured the vast majority of explainable within-person variation in social media use and that fine-tuning provided only limited value.

\subsection{Relationships Between Predictability and Behavioral Frequency}
To gain a better understanding of the factors driving individual differences in the predictability of social media use, we analyzed its associations with properties of the underlying training data. Specifically, we analyzed the association between predictability scores (person-level AUC scores generated by the various predictive models) and frequency measures as indicators of habitual behavior (total number of app sessions per person; absolute number of social media sessions per person; proportion of social media sessions relative to the total number of app sessions per person). Low correlations between predictability scores and behavioral frequency measures would indicate that (i) person-level model performance was not primarily driven by the availability of training data and (ii) that person-level predictability of behavioral sequences captures non-redundant information with respect to behavioral frequency measures. We used Pearson's correlation coefficients for all analyses.

We found negligible associations between person-level predictability scores and the total number of app sessions per person in the global  ($r_{lstm}$=-0.014; $r_{trans}$=0.058) and fine-tuned models  ($r_{lstm}$=-0.017; $r_{trans}$=0.029). As would be expected, the correlations were higher for person-specific models trained on person-level data ($r_{lstm}$=0.273, $r_{trans}$=0.322). Similarly, predictive performance scores were not related to the absolute number of social media sessions for global models  ($r_{lstm}$=-0.049, $r_{trans}$=0.004) and fine-tuned models ($r_{lstm}$=0.035, $r_{trans}$=0.003) but were somewhat related for person-specific models  ($r_{lstm}$=0.247; $r_{trans}$=0.258). The proportion of social media sessions relative to the total number of app sessions was weakly negatively related to model performance in global models  ($r_{lstm}$=-0.082; $r_{trans}$=-0.118) and fine-tuned models  ($r_{lstm}$=-0.052; $r_{trans}$=-0.092) but weakly positively related for person-specific models  ($r_{lstm}$=0.178; $r_{trans}$=0.124). These findings indicate that (i) differences in the predictability of social media use are not primarily driven by the availability of training data, even in the case of person-specific models, and (ii) they support the proposition that the predictability of social media use is indeed distinct from behavioral frequency habit measures.

\subsection{Relationships Between Predictability and Context Size}
\begin{figure}
  \centering
  \includegraphics[width=1\textwidth]{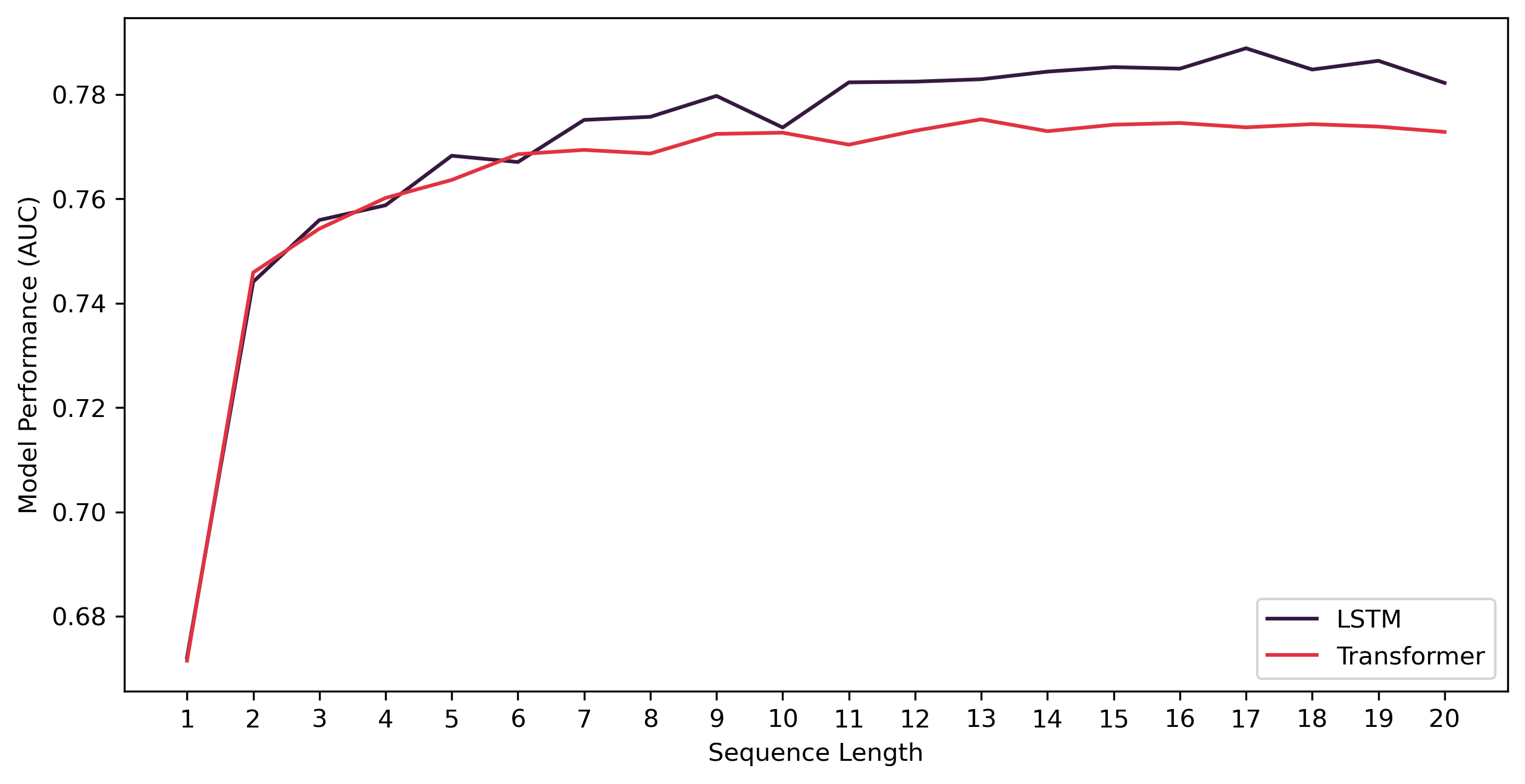}
  \caption[Global model performance as a function of sequence length]
  {Global model performance as a function of sequence length. While longer sequences were generally associated with higher predictive performance, the incremental performance gains decreased with sequence length.}
  \label{fig:habit_sequence_length}
\end{figure}

In order to examine the boundaries of the relevant behavioral context, we analyzed model performance as a function of input sequence length  (i.e., the number of preceding app sessions available to the model). For this purpose, we re-trained the global model for sequence lengths between one and twenty time steps and another model with 50 preceding app sessions, accounting for the possibility that far-removed behaviors would still contain useful information. The results show that longer sequences are generally associated with higher model performance, with diminishing incremental contributions of each additional time step. The LSTM model reached its best test performance at a sequence length of 17 with an AUC of 0.789, while the transformer model reached its best test performance at a sequence length of 13 with an AUC of 0.775. Models trained on sequences of 50 time steps did not perform better ($AUC_{lstm}$=0.787; $AUC_{trans}$=0.775).
The results also reveal that even extremely short sequences are predictive of subsequent social media use. For example, both models performed considerably above chance based on information about only the last preceding app session  ($AUC_{lstm}$=0.672; $AUC_{trans}$=0.671). Generally, the incremental performance improvement for each additional unit in sequence length leveled off dramatically for models trained on sequences of more than three to ten app sessions. For a visual presentation of the results, please refer to Figure \ref{fig:habit_sequence_length}.

\FloatBarrier

\section{Discussion}
The present paper is among the first to examine the predictability of social media use through the lens of media and technology habits. The results demonstrate that social media use is moderately to highly predictable from preceding smartphone user behaviors. This finding highlights the repetitive, predictable nature of social media habits and is aligned with past research on the predictability of more fine-grained online social behaviors such as active and passive social media use \cite{peters_context-aware_2023}, instant messaging \cite{peters_predicting_2022} or in-app action sequences \cite{liu_characterizing_2019}, as well as app usage in general \cite{baeza-yates_predicting_2015, huang_predicting_2012, natarajan_which_2013, parate_practical_2013, xia_deepapp_2020, xu_predicting_2020}.

Contributing specifically to the habits literature, our study highlights the importance and predictive validity of preceding action cues \cite{wood_habit_2017}, which have not been examined at this level of granularity in previous work. The results indicate that objectively captured preceding action cues contain considerable information about momentary social media use and that the sequential structure of app use matters above and beyond more general behavioral tendencies that have been studied in most previous work \cite[see][]{bayer_building_2022}). Relatedly, our analysis of the relationship between sequence lengths and predictive performance indicates that the relevant context window spans sequences of between roughly three and ten apps, as predictive performance did not increase substantially for longer sequences. While this finding is highly specific to the research setting and the predictive task at hand, it provides an intuitive way to distinguish between longer-term behavioral tendencies and preceding action cues. Behavioral tendencies are captured by the consistency of behavioral frequencies over time, whereas preceding action cues can be defined by their incremental predictive contribution of each prior step (e.g., app selection) with respect to a focal behavior (e.g., social media use).

Our results also address the question of the generalizability of social media habits across individuals. As the person-specific models did not generalize well across individuals, our results indicate that the models pick up on idiosyncratic habits \cite{gardner_review_2015, gardner_modelling_2018, lally_how_2010}. At the same time, the fact that person-specific models did not systematically outperform global models indicates that the global models not only represent general behavioral trends but simultaneously learn a broad range of idiosyncratic patterns, given sufficient training data. These findings raise the question as to whether individual differences in the predictability of social media use can be interpreted as an objective measure of habit strength, focusing on the predictably recurrent nature of habit. The fact that predictive performance scores were not highly correlated with frequency measures of social media use indicates that they capture a fundamentally different aspect of social media habits. While the interpretation of predictability as habit strength still needs to be substantiated by dedicated psychometric analyses, it could open up new opportunities for studying habits based on the wealth of objective behavioral data that is available through consumer electronics and online social platforms \cite{harari_using_2016, peters_big_2022}. 

More broadly, the current research represents a novel approach to the study of media and technology habits and affirms the potential of modeling sequential behavior for the future of habit research.  By utilizing objective records of smartphone use, our study circumvents these limitations of self-report methods \cite{boyle_systematic_2022} while retaining the ecological validity of studying individuals’ own real-world habits. Moreover, the focus on predictive models designed to represent temporal dependencies presents a new step in the identification of media and technology habits. This approach aligns with the broader shift in behavioral science research towards leveraging objectively trackable data and machine learning to uncover patterns that are not readily apparent through traditional methods \cite{lazer_computational_2020, peters_big_2022, stachl_personality_2020}.

Notably, our analytical framework can be extended to generate a wide range of features related to the predictability of behavior that are not examined in the current research. Aside from analyzing predictive performance purely based on preceding actions, it would be possible to analyze predictive performance with regard to other theoretically relevant constructs. For example, the per-person uptick in predictive performance after adding location data to the model could be interpreted as a “context-contingency score”, indicating the varying degrees to which people's behaviors are determined by their environments. Similarly, the degree of generalizability of person-specific models could be interpreted as an “idiosyncrasy score”, indicating the extent to which individuals exhibit unique behavioral associations. Relatedly, measures of model complexity  (in the absence of regularization) in person-specific models could serve as a measure of habit complexity. Future research should investigate the viability of these suggestions and perform dedicated psychometric analyses.

Finally, our approach may not be limited to the study of social media habits but could also be applied to other behavioral phenomena and theoretical perspectives, such as personality and person-situation transactions. For example, the Cognitive-Affective Personality System (CAPS) \cite{mischel_toward_1974, mischel_cognitive-affective_1995} defines personality not as a set of static traits but rather as a collection of individual differences in how people respond to varying situations. According to this theory, personality is understood through the distinctive patterns of thoughts, feelings, and behaviors that emerge as individuals interact with specific situational contexts. Our modeling approach would lend itself to examining this notion of personality based on extensive behavioral records. Finding a large predictive contribution of context features and relatively low generalizability across person-specific models would indicate the validity of CAPS with regard to online behaviors and would extend prior work on the behavioral consistency of smartphone use \cite{shaw_behavioral_2022}.

Taken together, the contributions of our work are theoretical, empirical, methodological, and practical in nature. First, our study focuses on habits as embedded in behavioral sequences, which is highlighted by the theoretical definition of habits but has not been studied sufficiently in past research. Our work derives a novel operationalization of media and technology habits (based on the concept of person-level predictability), which is theoretically and empirically distinct from behavioral frequency. Relatedly, we analyze and discuss app sequences from the perspective of preceding action cues - a component process of special interest to the habits literature. Second, our study utilizes a unique dataset of highly granular, objective, behavioral data that allows us to study social media habits with high ecological validity. Third, our study pioneers the use of sequential machine learning techniques for habit modeling. Finally, our research has practical implications as the ability to predict user behaviors allows for more effective mobile technologies and personalized user experiences. We discuss these implications in greater detail below.

\subsection{Limitations and Directions for Future Research}
Our work has several limitations that need to be addressed in future research. Firstly, while the overall volume of data was sufficient to train complex machine learning models, we faced data limitations for the person-specific models. The lack of training data per person is one potential explanation for the finding that person-specific models were outperformed by global models. The relatively low correlation between person-level AUC scores and the number of data points per person, however, suggests that differences in predictability are not merely an artifact introduced by varying degrees of data availability across individuals. Nonetheless, additional research is needed in order to fully rule out concerns related to data coverage. Mirroring the concerns about data coverage for training person-specific models, insufficient fine-tuning data could have decreased the effectiveness of the fine-tuning stage. Importantly, the limited effectiveness of person-specific modeling and fine-tuning could also be related to the decision to frame the predictive task as a binary classification task (i.e., collapsing various social media apps into the single category of “social media use”). This approach likely attenuated some of the idiosyncratic patterns that person-specific or fine-tuned models might have otherwise picked up on. Future research should therefore consider examining a more granular range of target behaviors.

Second, by condensing all social media use into a single category, our approach does not distinguish between different platforms, their unique affordances \cite{bucher_affordances_2018}, or the nature of the engagement (e.g., active versus passive use) \cite{escobar-viera_passive_2018, hemmings-jarrett_evaluation_2017, khan_social_2017}. While this decision was necessary to standardize the target variable across models, it leaves room for further exploration, as different social media platforms have distinct features and may engender different patterns of usage. For instance, the usage patterns on a visually oriented platform like Instagram may differ from text-based platforms like X or Reddit. Additionally, active use (such as posting or commenting) likely follows different habitual patterns compared to passive use (such as browsing without interacting) – and these differences may have implications for well-being \cite{verduyn_passive_2015}. Future research should take these distinctions into account and examine how different types of social media engagement impact user behavior and experience. This could lead to a more nuanced understanding of social media habits and their effects on individuals, potentially offering more targeted insights for the development of interventions for healthier technology use. 

Third, we did not analyze the relationships between person-level predictive performance and self-reported habit measures. Future research should alleviate this concern and also analyze the relationships between predictability scores and other variables known to be related to habitual social media use. However, as we maintain that the predictability of behavior captures an aspect of habits not sufficiently represented in self-report measures, we would not necessarily expect a high correlation. A better empirical test of criterion validity would be whether individual differences in the predictability of social media use show incremental explanatory power above and beyond existing self-report measures. Additionally, future work can compare how preceding sequences relate to other indirect indicators of habitual behavior (e.g., devaluation insensitivity). Overall, our approach is not intended to replace existing habit measures but to provide a complementary operationalization that highlights their embeddedness in repetitive behavioral sequences.

Fourth, we exclusively focused on preceding user behaviors as input features without incorporating temporal information beyond the order of the logged app sessions or other contextual features that could provide a more complete picture of social media usage patterns. While adding time-related features did not lead to improved predictive performance in exploratory analyses, truncating the sequences at the daily level did not lead to a decrease in model performance, indicating that the current approach is robust to changes in the processing of app sequences (please see SI F for additional results). Nonetheless, including other contextual data, such as time of day or location, could allow for a more explicit examination of the context-contingent nature of social media habits in future research. Similarly, examining the effects of technical cues (e.g., notifications) and data from other smartphone sensors could add nuance to our research. This approach could potentially improve the predictive performance of our models by identifying contextual conditions under which certain behaviors are more or less likely to occur. 

Fifth, our work derives a novel operationalization of media and technology habits that is grounded in the predictability of behavioral sequences and is aligned with traditional definitions of habits highlighting their repetitive, predictable character. While this approach is suitable to shed light on the importance of preceding action cues and the generalizability of social media habits, our work does not directly speak to the causal mechanisms explaining specific user behaviors or habit activation processes. As such, our exploratory approach to modeling habitual patterns should be extended in future research to directly tap into habit formation and activation processes – along with the potentially causal role of habitual behavior in daily life.

Relatedly, while we maintain that the predictability of behavioral sequences is an indicator of habitual behavior, we acknowledge that there are non-habitual drivers of smartphone user behaviors in general and social media use specifically. As past research has shown, deliberate goal-directed behavior can become automated over time and is intimately linked to habit formation \cite{wood_new_2007, wood_psychology_2016}. Given our framework, it is likely that the process of habit formation would correspond to increasing levels of predictability as goal-directed behaviors give way to habits through repetition. Future work should further examine the distinction between habitual and goal-directed social media use and identify objective markers associated with these different orientations. This could include a more detailed analysis of specific types of user behaviors. For instance, it is conceivable that passive social media use (content consumption) is more habit-driven than active use (social interaction), as the latter typically requires higher cognitive engagement and potentially more conscious decision-making. 

Finally, as habits may not be perfectly stable over time, changes in user behavior could limit the predictability of social media use. Such changes could be related to habit formation and habit cessation, or due to observer effects. In our framework, the presence of such changes and the resulting reduction in predictive performance would indicate a lower degree of habit strength (via reduced repetitiveness in sequence patterns). While this is not necessarily a weakness of our approach, future work could further examine how the predictability of social media use varies over longer time frames and how these fluctuations relate to habit formation and cessation.

\subsection{Practical Implications}
Aside from its theoretical implications, our work lays the foundations for more habit-centered approaches to user modeling. This shift could, in turn, lead to the development of new technologies and user experiences that are more aligned with users’ needs and preferences. For example, the ability to predict when and what type of app a user is likely to engage with can be utilized for prefetching content \cite{parate_practical_2013}. This means that apps could preload content before a user is likely to engage. This not only enhances the user experience by reducing loading times but can also help to manage network traffic more effectively by spreading content download over time. Importantly, while such applications can improve user experiences, they should be implemented in a way that does not nudge the user to perform unwanted habitual behavior. 

Another potential opportunity lies in adaptive user interfaces and notifications. Interfaces could adjust their layout and content to align with the user's predicted preferences and habits at different times. For example, apps that are likely to be used could be presented on the home screen for easy access. Similarly, knowing when users are most likely to engage with social media could help time notifications and alerts more effectively. As smartphone use can have negative effects on productivity \cite{derks_private_2021, duke_smartphone_2017, ward_brain_2017}, predictions of social media engagement could also be used to optimize the delivery timing of messages and avoid unwanted interruptions. A similar idea has been discussed as “bounded deferral” \cite{dingler_ill_2015, hutchison_understanding_2005}, where messages are held back while a user is busy, but only up to a maximum amount of time. 

For users who wish to reduce their social media engagement, predictive models can make interventions more personalized and effective. For example, it has been shown that introducing friction (e.g., a short wait period before opening a social media app) can be highly effective in reducing social media use \cite{gruning_directing_2023}. Such interventions could be dynamically adjusted based on ongoing or preceding user behavior, making them more relevant and effective. For instance, users might want to intervene in habit-driven use but not goal-directed use. 

Finally, a better understanding of social media habits can help policymakers develop user-centric policies to protect young users from the potential harms of excessive media consumption and promote digital well-being \cite{vanden_abeele_digital_2021}.

\subsection{Conclusion}
Taken together, our results indicate that social media use is predictable within and across individuals and that global models are capable of making accurate predictions, even at the within-person level. By utilizing objective behavioral time series data, employing advanced predictive models, and examining the predictability of social media use at the within-person level, our work highlights a new direction in habit research. Our findings shed light on the predictive validity of preceding action cues and the value of sequential information in predicting social media habits. In addition, our findings reveal individual differences in the predictability of social media use and raise the question as to whether they can be interpreted as a novel indicator of habit strength. Finally, our study not only enhances our understanding of media consumption behaviors but also paves the way for more data-driven strategies to manage everyday media and technology habits.

\newpage

\section*{Ethics approval}
The study was approved by the Stanford University IRB (Protocol \#48234). All methods were carried out in accordance with relevant guidelines and regulations.

\section*{Availability of data and materials}
The code used to generate the results is available on this paper’s OSF page (\href{https://osf.io/rkswe/}{https://osf.io/rkswe/}). The data needed to reproduce the analyses will be made available after peer review of the manuscript is complete. 

\section*{Competing interests}
The authors declare no potential conflicts of interest.

\section*{Author contributions}
HP: Conceptualization, Methodology, Software, Formal analysis, Visualization, Writing - Original Draft; JBB: Conceptualization, Writing - Review \& Editing; SCM: Conceptualization, Writing - Review \& Editing; YC: Data Curation, Software; SSV: Writing - Review \& Editing; GMH: Conceptualization, Resources, Writing - Review \& Editing, Supervision, Project administration, Funding acquisition

\section*{Acknowledgments}
This research was supported in part by the  Stanford Institute for Human-Centered Artificial Intelligence (HAI) with a Google Cloud Credit Award and a Stanford HAI Seed Grant. The authors would like to thank Katherine Roehrick and Serena Soh for their research assistance with this study and their role in data collection.

\newpage
\printbibliography

@article{liu_characterizing_2019,
	title = {Characterizing and {Forecasting} {User} {Engagement} with {In}-app {Action} {Graph}: {A} {Case} {Study} of {Snapchat}},
	shorttitle = {Characterizing and {Forecasting} {User} {Engagement} with {In}-app {Action} {Graph}},
	url = {http://arxiv.org/abs/1906.00355},
	urldate = {2021-07-27},
	journal = {arXiv:1906.00355 [cs, stat]},
	author = {Liu, Yozen and Shi, Xiaolin and Pierce, Lucas and Ren, Xiang},
	month = jun,
	year = {2019},
	note = {arXiv: 1906.00355},
}

@inproceedings{hemmings-jarrett_evaluation_2017,
	title = {Evaluation of {User} {Engagement} on {Social} {Media} to {Leverage} {Active} and {Passive} {Communication}},
	doi = {10.1109/IEEE.ICCC.2017.24},
	booktitle = {2017 {IEEE} {International} {Conference} on {Cognitive} {Computing} ({ICCC})},
	author = {Hemmings-Jarrett, Kimberley and Jarrett, Julian and Blake, M. Brian},
	month = jun,
	year = {2017},
	pages = {132--135},
}

@article{khan_social_2017,
	title = {Social media engagement: {What} motivates user participation and consumption on {YouTube}?},
	volume = {66},
	issn = {07475632},
	shorttitle = {Social media engagement},
	url = {https://linkinghub.elsevier.com/retrieve/pii/S0747563216306513},
	doi = {10.1016/j.chb.2016.09.024},
	language = {en},
	urldate = {2021-07-08},
	journal = {Computers in Human Behavior},
	author = {Khan, M. Laeeq},
	month = jan,
	year = {2017},
	pages = {236--247},
}

@article{escobar-viera_passive_2018,
	title = {Passive and {Active} {Social} {Media} {Use} and {Depressive} {Symptoms} {Among} {United} {States} {Adults}},
	volume = {21},
	issn = {2152-2715, 2152-2723},
	url = {http://www.liebertpub.com/doi/10.1089/cyber.2017.0668},
	doi = {10.1089/cyber.2017.0668},
	language = {en},
	number = {7},
	urldate = {2021-07-08},
	journal = {Cyberpsychology, Behavior, and Social Networking},
	author = {Escobar-Viera, César G. and Shensa, Ariel and Bowman, Nicholas D. and Sidani, Jaime E. and Knight, Jennifer and James, A. Everette and Primack, Brian A.},
	month = jul,
	year = {2018},
	pages = {437--443},
}

@article{xia_deepapp_2020,
	title = {{DeepApp}: {Predicting} {Personalized} {Smartphone} {App} {Usage} via {Context}-{Aware} {Multi}-{Task} {Learning}},
	volume = {11},
	issn = {2157-6904, 2157-6912},
	shorttitle = {{DeepApp}},
	url = {https://dl.acm.org/doi/10.1145/3408325},
	doi = {10.1145/3408325},
	language = {en},
	number = {6},
	urldate = {2022-03-02},
	journal = {ACM Transactions on Intelligent Systems and Technology},
	author = {Xia, Tong and Li, Yong and Feng, Jie and Jin, Depeng and Zhang, Qing and Luo, Hengliang and Liao, Qingmin},
	month = nov,
	year = {2020},
	pages = {1--12},
}

@article{schmidhuber_deep_2015,
	title = {Deep learning in neural networks: {An} overview},
	volume = {61},
	issn = {08936080},
	shorttitle = {Deep learning in neural networks},
	url = {https://linkinghub.elsevier.com/retrieve/pii/S0893608014002135},
	doi = {10.1016/j.neunet.2014.09.003},
	language = {en},
	urldate = {2022-01-28},
	journal = {Neural Networks},
	author = {Schmidhuber, Jürgen},
	month = jan,
	year = {2015},
	pages = {85--117},
}

@article{hochreiter_long_1997,
	title = {Long {Short}-{Term} {Memory}},
	volume = {9},
	issn = {0899-7667},
	url = {https://doi.org/10.1162/neco.1997.9.8.1735},
	doi = {10.1162/neco.1997.9.8.1735},
	number = {8},
	urldate = {2022-01-25},
	journal = {Neural Computation},
	author = {Hochreiter, Sepp and Schmidhuber, Jürgen},
	month = nov,
	year = {1997},
	pages = {1735--1780},
}

@incollection{hutchison_understanding_2005,
	address = {Berlin, Heidelberg},
	title = {Understanding {Context} {Before} {Using} {It}},
	volume = {3554},
	isbn = {978-3-540-26924-3 978-3-540-31890-3},
	url = {http://link.springer.com/10.1007/11508373_3},
	language = {en},
	urldate = {2021-12-14},
	booktitle = {Modeling and {Using} {Context}},
	publisher = {Springer Berlin Heidelberg},
	author = {Bazire, Mary and Brézillon, Patrick},
	editor = {Hutchison, David and Kanade, Takeo and Kittler, Josef and Kleinberg, Jon M. and Mattern, Friedemann and Mitchell, John C. and Naor, Moni and Nierstrasz, Oscar and Pandu Rangan, C. and Steffen, Bernhard and Sudan, Madhu and Terzopoulos, Demetri and Tygar, Dough and Vardi, Moshe Y. and Weikum, Gerhard and Dey, Anind and Kokinov, Boicho and Leake, David and Turner, Roy},
	year = {2005},
	doi = {10.1007/11508373_3},
	note = {Series Title: Lecture Notes in Computer Science},
	pages = {29--40},
}

@article{wood_new_2007,
	title = {A new look at habits and the habit-goal interface.},
	volume = {114},
	issn = {1939-1471},
	url = {https://psycnet-apa-org.ezproxy.cul.columbia.edu/fulltext/2007-13558-001.pdf},
	doi = {10.1037/0033-295X.114.4.843},
	number = {4},
	urldate = {2021-11-09},
	journal = {Psychological Review},
	author = {Wood, Wendy and Neal, David T.},
	year = {2007},
	note = {Publisher: US: American Psychological Association},
	pages = {843},
}

@article{naab_habitual_2016,
	title = {Habitual {Initiation} of {Media} {Use} and a {Response}-{Frequency} {Measure} for {Its} {Examination}},
	volume = {19},
	issn = {1521-3269, 1532-785X},
	url = {http://www.tandfonline.com/doi/full/10.1080/15213269.2014.951055},
	doi = {10.1080/15213269.2014.951055},
	language = {en},
	number = {1},
	urldate = {2021-11-09},
	journal = {Media Psychology},
	author = {Naab, Teresa K. and Schnauber, Anna},
	month = jan,
	year = {2016},
	pages = {126--155},
}

@article{schnauber-stockmann_process_2019,
	title = {The process of forming a mobile media habit: results of a longitudinal study in a real-world setting},
	volume = {22},
	issn = {1521-3269, 1532-785X},
	shorttitle = {The process of forming a mobile media habit},
	url = {https://www.tandfonline.com/doi/full/10.1080/15213269.2018.1513850},
	doi = {10.1080/15213269.2018.1513850},
	language = {en},
	number = {5},
	urldate = {2021-11-09},
	journal = {Media Psychology},
	author = {Schnauber-Stockmann, Anna and Naab, Teresa K.},
	month = sep,
	year = {2019},
	pages = {714--742},
}

@incollection{verplanken_technology_2018,
	address = {Cham},
	title = {Technology {Habits}: {Progress}, {Problems}, and {Prospects}},
	isbn = {978-3-319-97528-3 978-3-319-97529-0},
	shorttitle = {Technology {Habits}},
	url = {http://link.springer.com/10.1007/978-3-319-97529-0_7},
	language = {en},
	urldate = {2021-10-28},
	booktitle = {The {Psychology} of {Habit}},
	publisher = {Springer International Publishing},
	author = {Bayer, Joseph B. and LaRose, Robert},
	editor = {Verplanken, Bas},
	year = {2018},
	doi = {10.1007/978-3-319-97529-0_7},
	pages = {111--130},
}

@article{anderson_habits_2021,
	title = {Habits and the electronic herd: {The} psychology behind social media’s successes and failures},
	volume = {4},
	issn = {2476-1281},
	shorttitle = {Habits and the electronic herd},
	url = {https://onlinelibrary.wiley.com/doi/abs/10.1002/arcp.1063},
	doi = {10.1002/arcp.1063},
	language = {en},
	number = {1},
	urldate = {2021-10-28},
	journal = {Consumer Psychology Review},
	author = {Anderson, Ian A. and Wood, Wendy},
	year = {2021},
	note = {\_eprint: https://onlinelibrary.wiley.com/doi/pdf/10.1002/arcp.1063},
	pages = {83--99},
}

@incollection{gardner_habit_2019,
	title = {Habit {Formation} and {Behavior} {Change}},
	isbn = {978-0-19-023655-7},
	url = {https://oxfordre.com/psychology/view/10.1093/acrefore/9780190236557.001.0001/acrefore-9780190236557-e-129},
	language = {en},
	urldate = {2021-10-28},
	booktitle = {Oxford {Research} {Encyclopedia} of {Psychology}},
	publisher = {Oxford University Press},
	author = {Gardner, Benjamin and Rebar, Amanda L.},
	collaborator = {Gardner, Benjamin and Rebar, Amanda L.},
	month = apr,
	year = {2019},
	doi = {10.1093/acrefore/9780190236557.013.129},
}

@article{bayer_texting_2012,
	title = {Texting while driving on automatic: {Considering} the frequency-independent side of habit},
	volume = {28},
	issn = {0747-5632},
	shorttitle = {Texting while driving on automatic},
	url = {https://www.sciencedirect.com/science/article/pii/S0747563212001628},
	doi = {10.1016/j.chb.2012.06.012},
	language = {en},
	number = {6},
	urldate = {2021-10-28},
	journal = {Computers in Human Behavior},
	author = {Bayer, Joseph B. and Campbell, Scott W.},
	month = nov,
	year = {2012},
	pages = {2083--2090},
}

@inproceedings{siami-namini_performance_2019,
	title = {The {Performance} of {LSTM} and {BiLSTM} in {Forecasting} {Time} {Series}},
	doi = {10.1109/BigData47090.2019.9005997},
	booktitle = {2019 {IEEE} {International} {Conference} on {Big} {Data} ({Big} {Data})},
	author = {Siami-Namini, Sima and Tavakoli, Neda and Namin, Akbar Siami},
	month = dec,
	year = {2019},
	pages = {3285--3292},
}

@inproceedings{graves_hybrid_2013,
	title = {Hybrid speech recognition with {Deep} {Bidirectional} {LSTM}},
	doi = {10.1109/ASRU.2013.6707742},
	booktitle = {2013 {IEEE} {Workshop} on {Automatic} {Speech} {Recognition} and {Understanding}},
	author = {Graves, Alex and Jaitly, Navdeep and Mohamed, Abdel-rahman},
	month = dec,
	year = {2013},
	pages = {273--278},
}

@inproceedings{han_ese_2017,
	address = {New York, NY, USA},
	series = {{FPGA} '17},
	title = {{ESE}: {Efficient} {Speech} {Recognition} {Engine} with {Sparse} {LSTM} on {FPGA}},
	isbn = {978-1-4503-4354-1},
	shorttitle = {{ESE}},
	url = {https://doi.org/10.1145/3020078.3021745},
	doi = {10.1145/3020078.3021745},
	urldate = {2022-04-21},
	booktitle = {Proceedings of the 2017 {ACM}/{SIGDA} {International} {Symposium} on {Field}-{Programmable} {Gate} {Arrays}},
	publisher = {Association for Computing Machinery},
	author = {Han, Song and Kang, Junlong and Mao, Huizi and Hu, Yiming and Li, Xin and Li, Yubin and Xie, Dongliang and Luo, Hong and Yao, Song and Wang, Yu and Yang, Huazhong and Dally, William (Bill) J.},
	month = feb,
	year = {2017},
	pages = {75--84},
}

@incollection{tokunaga_media_2020,
	title = {Media {Use} as {Habit}},
	isbn = {978-1-119-01107-1},
	url = {https://onlinelibrary.wiley.com/doi/abs/10.1002/9781119011071.iemp0102},
	language = {en},
	urldate = {2022-05-01},
	booktitle = {The {International} {Encyclopedia} of {Media} {Psychology}},
	publisher = {John Wiley \& Sons, Ltd},
	author = {Tokunaga, Robert S.},
	year = {2020},
	doi = {10.1002/9781119011071.iemp0102},
	note = {\_eprint: https://onlinelibrary.wiley.com/doi/pdf/10.1002/9781119011071.iemp0102},
	pages = {1--5},
}

@inproceedings{falkner_bohb_2018,
	title = {{BOHB}: {Robust} and {Efficient} {Hyperparameter} {Optimization} at {Scale}},
	shorttitle = {{BOHB}},
	url = {https://proceedings.mlr.press/v80/falkner18a.html},
	language = {en},
	urldate = {2022-05-05},
	booktitle = {Proceedings of the 35th {International} {Conference} on {Machine} {Learning}},
	publisher = {PMLR},
	author = {Falkner, Stefan and Klein, Aaron and Hutter, Frank},
	month = jul,
	year = {2018},
	note = {ISSN: 2640-3498},
	pages = {1437--1446},
}

@incollection{peters_big_2022,
	address = {Washington, DC, US},
	title = {The {Big} {Data} toolkit for psychologists: {Data} sources and methodologies},
	isbn = {978-1-4338-3626-8 978-1-4338-3847-7},
	shorttitle = {The {Big} {Data} toolkit for psychologists},
	booktitle = {The psychology of technology: {Social} science research in the age of {Big} {Data}},
	publisher = {American Psychological Association},
	author = {Peters, Heinrich and Marrero, Zachariah and Gosling, Samuel D.},
	year = {2022},
	doi = {10.1037/0000290-004},
	pages = {87--124},
}

@article{bayer_connection_2016,
	title = {Connection {Cues}: {Activating} the {Norms} and {Habits} of {Social} {Connectedness}},
	volume = {26},
	issn = {1468-2885},
	shorttitle = {Connection {Cues}},
	url = {https://onlinelibrary.wiley.com/doi/abs/10.1111/comt.12090},
	doi = {10.1111/comt.12090},
	language = {en},
	number = {2},
	urldate = {2023-01-30},
	journal = {Communication Theory},
	author = {Bayer, Joseph B. and Campbell, Scott W. and Ling, Rich},
	year = {2016},
	note = {\_eprint: https://onlinelibrary.wiley.com/doi/pdf/10.1111/comt.12090},
	pages = {128--149},
}

@book{bucher_affordances_2018,
	title = {The {Affordances} of {Social} {Media} {Platforms}},
	isbn = {978-1-4129-6229-2},
	url = {https://dare.uva.nl/search?identifier=149a9089-49a4-454c-b935-a6ea7f2d8986},
	language = {en},
	urldate = {2024-02-03},
	publisher = {Sage Publications},
	author = {Bucher, T. and Helmond, A.},
	year = {2018},
}

@article{bayer_building_2022,
	title = {Building and {Breaking} {Social} {Media} {Habits}},
	volume = {45},
	doi = {10.1016/j.copsyc.2022.101303},
	journal = {Current Opinion in Psychology},
	author = {Bayer, Joseph and Anderson, Ian and Tokunaga, Robert},
	month = jan,
	year = {2022},
	pages = {101303},
}

@article{amaya_neurobiology_2018,
	series = {Habits and {Skills}},
	title = {Neurobiology of habit formation},
	volume = {20},
	issn = {2352-1546},
	url = {https://www.sciencedirect.com/science/article/pii/S235215461730089X},
	doi = {10.1016/j.cobeha.2018.01.003},
	urldate = {2024-06-22},
	journal = {Current Opinion in Behavioral Sciences},
	author = {Amaya, Kenneth A and Smith, Kyle S},
	month = apr,
	year = {2018},
	pages = {145--152},
}

@article{orbell_automatic_2010,
	title = {The automatic component of habit in health behavior: {Habit} as cue-contingent automaticity},
	volume = {29},
	issn = {1930-7810},
	shorttitle = {The automatic component of habit in health behavior},
	doi = {10.1037/a0019596},
	number = {4},
	journal = {Health Psychology},
	author = {Orbell, Sheina and Verplanken, Bas},
	year = {2010},
	note = {Place: US
Publisher: American Psychological Association},
	pages = {374--383},
}

@article{gardner_habit_2012,
	title = {Habit as automaticity, not frequency},
	volume = {14},
	number = {2},
	journal = {European Health Psychologist},
	author = {Gardner, Benjamin},
	month = jun,
	year = {2012},
	pages = {32--36},
}

@article{docherty_facebooks_2020,
	title = {Facebook’s {Ideal} {User}: {Healthy} {Habits}, {Social} {Capital}, and the {Politics} of {Well}-{Being} {Online}},
	volume = {6},
	shorttitle = {Facebook’s {Ideal} {User}},
	doi = {10.1177/2056305120915606},
	journal = {Social Media + Society},
	author = {Docherty, Niall},
	month = apr,
	year = {2020},
	pages = {205630512091560},
}

@article{docherty_digital_2021,
	title = {Digital {Self}-{Control} and the {Neoliberalization} of {Social} {Media} {Well}-{Being}},
	volume = {15},
	copyright = {The  International Journal of Communication  is an academic journal. As such, it is dedicated to the open exchange of information. For this reason, IJoC is freely available to individuals and institutions. Copies of this journal or articles in this journal may be distributed for research or educational purposes free of charge and without permission. However, commercial use of the IJoC website or the articles contained herein is expressly prohibited without the written consent of the editor. Authors who publish in The  International Journal of Communication  will release their articles under the   Creative Commons Attribution Non-Commercial No Derivatives (by-nc-nd) license  . This license allows anyone to copy and distribute the article for non-commercial purposes provided that appropriate attribution is given. For details of the rights authors grants users of their work, see the  "human-readable summary" of the license , with a link to the full license. (Note that "you" refers to a user, not an author, in the summary.) This journal utilizes the  LOCKSS system to create a distributed archiving system among participating libraries and permits those libraries to create permanent archives of the journal for purposes of preservation and restoration. The publisher perpetually authorizes participants in the LOCKSS system to archive and restore our publication through the LOCKSS System for the benefit of all LOCKSS System participants. Specifically participating libraries may:  Collect and preserve currently accessible materials;  Use material consistent with original license terms;  Provide copies to other LOCKSS appliances for purposes of audit and repair.        Fair Use The U.S. Copyright Act of 1976 specifies, in Section 107, the terms of the Fair Use exception: Notwithstanding the provisions of sections 106 and 106A, the fair use of a copyrighted work, including such use by reproduction in copies or phonorecords or by any other means specified by that section, for purposes such as criticism, comment, news reporting, teaching (including multiple copies for classroom use), scholarship, or research, is not an infringement of copyright. In determining whether the use made of a work in any particular case is a fair use the factors to be considered shall include:  the purpose and character of the use, including whether such use is of a commercial nature or is for nonprofit educational purposes;  the nature of the copyrighted work;  the amount and substantiality of the portion used in relation to the copyrighted work as a whole; \&amp;  the effect of the use upon the potential market for or value of the copyrighted work.   The fact that a work is unpublished shall not itself bar a finding of fair use if such finding is made upon consideration of all the above factors. In accord with these provisions, the  International Journal of Communication  believes in the vigorous assertion and defense of Fair Use by scholars engaged in academic research, teaching and non-commercial publishing. Thus, we view the inclusion of “quotations” from existing print, visual, audio and audio-visual texts to be appropriate examples of Fair Use, as are reproductions of visual images for the purpose of scholarly analysis. We encourage authors to obtain appropriate permissions to use materials originally produced by others, but do not require such permissions as long as the usage of such materials falls within the boundaries of Fair Use.  The  International Journal of Communication  encourages authors to employ fair use in their scholarly publishing wherever appropriate. Fair use is the right to use unlicensed copyrighted material (whether it is text, images, audio-visual, or other) in your own work, in some circumstances. We consult the  Code of Best Practices in Fair Use for Scholarly Research in Communication , created by the International Communication Association and endorsed by the National Communication Association, and you should too. If you have any questions about whether fair use applies to your uses of copyrighted material (whether it is text, images, audio-visual, or other) in your scholarship, simply include your rationale, grounded in the Best Practices, as a supplementary document with your submission.},
	issn = {1932-8036},
	url = {https://ijoc.org/index.php/ijoc/article/view/17721},
	language = {en},
	number = {0},
	urldate = {2024-06-21},
	journal = {International Journal of Communication},
	author = {Docherty, Niall},
	month = aug,
	year = {2021},
	note = {Number: 0},
	pages = {20},
}

@incollection{larose_social_2011,
	title = {Social {Networking}: {Addictive}, {Compulsive}, {Problematic} or {Just} {Another} {Media} {Habit}?},
	isbn = {978-0-415-80180-5},
	shorttitle = {Social {Networking}},
	booktitle = {A {Networked} {Self}: {Identity}, {Community}, and {Culture} on {Social} {Network} {Sites}},
	author = {Larose, Robert and Kim, Jung-Hyun and Peng, Wei},
	month = jan,
	year = {2011},
	note = {Journal Abbreviation: A Networked Self: Identity, Community, and Culture on Social Network Sites},
	pages = {59--81},
}

@article{buyalskaya_what_2023,
	title = {What can machine learning teach us about habit formation? {Evidence} from exercise and hygiene},
	volume = {120},
	issn = {0027-8424, 1091-6490},
	shorttitle = {What can machine learning teach us about habit formation?},
	url = {https://pnas.org/doi/10.1073/pnas.2216115120},
	doi = {10.1073/pnas.2216115120},
	language = {en},
	number = {17},
	urldate = {2024-06-03},
	journal = {Proceedings of the National Academy of Sciences},
	author = {Buyalskaya, Anastasia and Ho, Hung and Milkman, Katherine L. and Li, Xiaomin and Duckworth, Angela L. and Camerer, Colin},
	month = apr,
	year = {2023},
	pages = {e2216115120},
}

@article{venkatesh_consumer_2012,
	title = {Consumer {Acceptance} and {Use} of {Information} {Technology}: {Extending} the {Unified} {Theory} of {Acceptance} and {Use} of {Technology}},
	volume = {36},
	issn = {0276-7783},
	shorttitle = {Consumer {Acceptance} and {Use} of {Information} {Technology}},
	url = {https://www.jstor.org.ezproxy.cul.columbia.edu/stable/41410412},
	doi = {10.2307/41410412},
	number = {1},
	urldate = {2024-06-03},
	journal = {MIS Quarterly},
	author = {Venkatesh, Viswanath and Thong, James Y. L. and Xu, Xin},
	year = {2012},
	note = {Publisher: Management Information Systems Research Center, University of Minnesota},
	pages = {157--178},
}

@inproceedings{devlin_bert_2019,
	address = {Minneapolis, MN},
	title = {{BERT}: {Pre}-training of {Deep} {Bidirectional} {Transformers} for {Language} {Understanding}},
	shorttitle = {{BERT}},
	url = {https://aclanthology.org/N19-1423},
	doi = {10.18653/v1/N19-1423},
	urldate = {2023-09-05},
	booktitle = {Proceedings of the 2019 {Conference} of the {North} {American} {Chapter} of the {Association} for {Computational} {Linguistics}: {Human} {Language} {Technologies}},
	publisher = {Association for Computational Linguistics},
	author = {Devlin, Jacob and Chang, Ming-Wei and Lee, Kenton and Toutanova, Kristina},
	month = jun,
	year = {2019},
	pages = {4171--4186},
}

@article{mendelsohn_creatures_2019,
	title = {Creatures of {Habit}: {The} {Neuroscience} of {Habit} and {Purposeful} {Behavior}},
	volume = {85},
	issn = {0006-3223},
	shorttitle = {Creatures of {Habit}},
	url = {https://www.ncbi.nlm.nih.gov/pmc/articles/PMC6701929/},
	doi = {10.1016/j.biopsych.2019.03.978},
	number = {11},
	urldate = {2024-04-03},
	journal = {Biological psychiatry},
	author = {Mendelsohn, Alana I.},
	month = jun,
	year = {2019},
	pmid = {31122343},
	pmcid = {PMC6701929},
	pages = {49--51},
}

@article{nebe_characterizing_2024,
	title = {Characterizing {Human} {Habits} in the {Lab}},
	volume = {10},
	issn = {2474-7394},
	url = {https://doi.org/10.1525/collabra.92949},
	doi = {10.1525/collabra.92949},
	number = {1},
	urldate = {2024-04-03},
	journal = {Collabra: Psychology},
	author = {Nebe, Stephan and Kretzschmar, André and Brandt, Maike C. and Tobler, Philippe N.},
	month = feb,
	year = {2024},
	pages = {92949},
}

@inproceedings{peters_predicting_2022,
	address = {Philadelphia, PA},
	title = {Predicting {Attentiveness} and {Responsiveness} in {Instant} {Messaging} -  {Evidence} from a {Large} {Online} {Social} {Platform}},
	booktitle = {6th-annual {Psychology} of {Technology} {Conference}},
	author = {Peters, Heinrich and Dotsch, Ron and Triguero Roura, Mireia and Liu, Yozen and Matz, Sandra C. and Bos, Maarten W.},
	year = {2022},
}

@article{shaw_behavioral_2022,
	title = {Behavioral {Consistency} in the {Digital} {Age}},
	volume = {33},
	issn = {0956-7976},
	url = {https://doi.org/10.1177/09567976211040491},
	doi = {10.1177/09567976211040491},
	language = {en},
	number = {3},
	urldate = {2024-04-01},
	journal = {Psychological Science},
	author = {Shaw, Heather and Taylor, Paul J. and Ellis, David A. and Conchie, Stacey M.},
	month = mar,
	year = {2022},
	note = {Publisher: SAGE Publications Inc},
	pages = {364--370},
}

@article{bayer_social_2020,
	title = {Social {Media} {Elements}, {Ecologies}, and {Effects}},
	volume = {71},
	issn = {0066-4308, 1545-2085},
	url = {https://www.annualreviews.org/doi/10.1146/annurev-psych-010419-050944},
	doi = {10.1146/annurev-psych-010419-050944},
	language = {en},
	number = {1},
	urldate = {2024-04-01},
	journal = {Annual Review of Psychology},
	author = {Bayer, Joseph B. and Triệu, Penny and Ellison, Nicole B.},
	month = jan,
	year = {2020},
	pages = {471--497},
}

@article{roffarello_understanding_2021,
	title = {Understanding and {Streamlining} {App} {Switching} {Experiences} in {Mobile} {Interaction}},
	volume = {158},
	doi = {10.1016/j.ijhcs.2021.102735},
	number = {2},
	journal = {International Journal of Human-Computer Studies},
	author = {Roffarello, Alberto and Russis, Luigi},
	month = nov,
	year = {2021},
	pages = {102735},
}

@article{vaid_variation_2024,
	title = {Variation in social media sensitivity across people and contexts},
	volume = {14},
	issn = {2045-2322},
	doi = {10.1038/s41598-024-55064-y},
	language = {eng},
	number = {1},
	journal = {Scientific Reports},
	author = {Vaid, Sumer S. and Kroencke, Lara and Roshanaei, Mahnaz and Talaifar, Sanaz and Hancock, Jeffrey T. and Back, Mitja D. and Gosling, Samuel D. and Ram, Nilam and Harari, Gabriella M.},
	month = mar,
	year = {2024},
	pmid = {38503817},
	pmcid = {PMC10951328},
	pages = {6571},
}

@article{peters_context-aware_2023,
	title = {Context-{Aware} {Prediction} of {User} {Engagement} on {Online} {Social} {Platforms}},
	url = {http://arxiv.org/abs/2310.14533},
	urldate = {2023-10-24},
	journal = {arXiv:2310.14533 [cs]},
	author = {Peters, Heinrich and Liu, Yozen and Barbieri, Francesco and Baten, Raiyan A. and Matz, Sandra C. and Bos, Maarten W.},
	month = oct,
	year = {2023},
	note = {arXiv:2310.14533 [cs]},
}

@article{kingma_adam_2017,
	title = {Adam: {A} {Method} for {Stochastic} {Optimization}},
	shorttitle = {Adam},
	url = {http://arxiv.org/abs/1412.6980},
	doi = {10.48550/arXiv.1412.6980},
	urldate = {2024-03-15},
	journal = {arXiv:1412.6980 [cs]},
	author = {Kingma, Diederik P. and Ba, Jimmy},
	month = jan,
	year = {2017},
	note = {arXiv:1412.6980 [cs]},
}

@inproceedings{dingler_ill_2015,
	address = {New York, NY},
	series = {{MobileHCI} '15},
	title = {I'll be there for you: {Quantifying} {Attentiveness} towards {Mobile} {Messaging}},
	isbn = {978-1-4503-3652-9},
	shorttitle = {I'll be there for you},
	url = {https://doi.org/10.1145/2785830.2785840},
	doi = {10.1145/2785830.2785840},
	urldate = {2023-01-04},
	booktitle = {Proceedings of the 17th {International} {Conference} on {Human}-{Computer} {Interaction} with {Mobile} {Devices} and {Services}},
	publisher = {ACM},
	author = {Dingler, Tilman and Pielot, Martin},
	month = aug,
	year = {2015},
	pages = {1--5},
}

@inproceedings{abadi_tensorflow_2016,
	address = {Savannah, GA},
	series = {{OSDI}'16},
	title = {{TensorFlow}: a system for large-scale machine learning},
	isbn = {978-1-931971-33-1},
	shorttitle = {{TensorFlow}},
	urldate = {2024-03-25},
	booktitle = {Proceedings of the 12th {USENIX} conference on {Operating} {Systems} {Design} and {Implementation}},
	publisher = {USENIX Association},
	author = {Abadi, Martín and Barham, Paul and Chen, Jianmin and Chen, Zhifeng and Davis, Andy and Dean, Jeffrey and Devin, Matthieu and Ghemawat, Sanjay and Irving, Geoffrey and Isard, Michael and Kudlur, Manjunath and Levenberg, Josh and Monga, Rajat and Moore, Sherry and Murray, Derek G. and Steiner, Benoit and Tucker, Paul and Vasudevan, Vijay and Warden, Pete and Wicke, Martin and Yu, Yuan and Zheng, Xiaoqiang},
	month = nov,
	year = {2016},
	pages = {265--283},
}

@article{ram_binding_2023,
	title = {Binding the {Person}-{Specific} {Approach} to {Modern} {AI} in the {Human} {Screenome} {Project}: {Moving} past {Generalizability} to {Transferability}},
	issn = {0027-3171},
	shorttitle = {Binding the {Person}-{Specific} {Approach} to {Modern} {AI} in the {Human} {Screenome} {Project}},
	url = {https://doi.org/10.1080/00273171.2023.2229305},
	doi = {10.1080/00273171.2023.2229305},
	urldate = {2023-12-14},
	journal = {Multivariate Behavioral Research},
	author = {Ram, Nilam and Haber, Nick and Robinson, Thomas N. and Reeves, Byron},
	year = {2023},
	pmid = {37439508},
	note = {Publisher: Routledge
\_eprint: https://doi.org/10.1080/00273171.2023.2229305},
	pages = {1--9},
}

@incollection{hamaker_idiographic_2009,
	address = {New York, NY, US},
	title = {Idiographic data analysis: {Quantitative} methods—from simple to advanced},
	isbn = {978-0-387-95921-4 978-0-387-95922-1},
	shorttitle = {Idiographic data analysis},
	booktitle = {Dynamic process methodology in the social and developmental sciences},
	publisher = {Springer Science + Business Media},
	author = {Hamaker, Ellen L. and Dolan, Conor V.},
	editor = {Valsiner, J. and Moleenar, P. C. M. and Lyra, M. C. D. P. and Chaudhary, N.},
	year = {2009},
	doi = {10.1007/978-0-387-95922-1_9},
	pages = {191--216},
}

@inproceedings{natarajan_which_2013,
	address = {New York, NY, USA},
	series = {{RecSys} '13},
	title = {Which app will you use next? collaborative filtering with interactional context},
	isbn = {978-1-4503-2409-0},
	shorttitle = {Which app will you use next?},
	url = {https://dl.acm.org/doi/10.1145/2507157.2507186},
	doi = {10.1145/2507157.2507186},
	urldate = {2024-01-31},
	booktitle = {Proceedings of the 7th {ACM} conference on {Recommender} systems},
	publisher = {ACM},
	author = {Natarajan, Nagarajan and Shin, Donghyuk and Dhillon, Inderjit S.},
	month = oct,
	year = {2013},
	pages = {201--208},
}

@inproceedings{parate_practical_2013,
	address = {New York, NY, USA},
	series = {{UbiComp} '13},
	title = {Practical prediction and prefetch for faster access to applications on mobile phones},
	isbn = {978-1-4503-1770-2},
	url = {https://dl.acm.org/doi/10.1145/2493432.2493490},
	doi = {10.1145/2493432.2493490},
	urldate = {2024-01-31},
	booktitle = {Proceedings of the 2013 {ACM} international joint conference on {Pervasive} and ubiquitous computing},
	publisher = {ACM},
	author = {Parate, Abhinav and Böhmer, Matthias and Chu, David and Ganesan, Deepak and Marlin, Benjamin M.},
	month = sep,
	year = {2013},
	pages = {275--284},
}

@inproceedings{huang_predicting_2012,
	address = {New York, NY, USA},
	series = {{UbiComp} '12},
	title = {Predicting mobile application usage using contextual information},
	isbn = {978-1-4503-1224-0},
	url = {https://dl.acm.org/doi/10.1145/2370216.2370442},
	doi = {10.1145/2370216.2370442},
	urldate = {2024-01-31},
	booktitle = {Proceedings of the 2012 {ACM} {Conference} on {Ubiquitous} {Computing}},
	publisher = {ACM},
	author = {Huang, Ke and Zhang, Chunhui and Ma, Xiaoxiao and Chen, Guanling},
	month = sep,
	year = {2012},
	pages = {1059--1065},
}

@inproceedings{baeza-yates_predicting_2015,
	address = {New York, NY, USA},
	series = {{WSDM} '15},
	title = {Predicting {The} {Next} {App} {That} {You} {Are} {Going} {To} {Use}},
	isbn = {978-1-4503-3317-7},
	url = {https://dl.acm.org/doi/10.1145/2684822.2685302},
	doi = {10.1145/2684822.2685302},
	urldate = {2024-01-31},
	booktitle = {Proceedings of the {Eighth} {ACM} {International} {Conference} on {Web} {Search} and {Data} {Mining}},
	publisher = {ACM},
	author = {Baeza-Yates, Ricardo and Jiang, Di and Silvestri, Fabrizio and Harrison, Beverly},
	month = feb,
	year = {2015},
	pages = {285--294},
}

@inproceedings{vaswani_attention_2017,
	address = {Red Hook, NY},
	series = {{NIPS}'17},
	title = {Attention is all you need},
	isbn = {978-1-5108-6096-4},
	urldate = {2024-01-29},
	booktitle = {Proceedings of the 31st {International} {Conference} on {Neural} {Information} {Processing} {Systems}},
	publisher = {Curran Associates},
	author = {Vaswani, Ashish and Shazeer, Noam and Parmar, Niki and Uszkoreit, Jakob and Jones, Llion and Gomez, Aidan N. and Kaiser, Łukasz and Polosukhin, Illia},
	month = dec,
	year = {2017},
	pages = {6000--6010},
}

@article{vanden_abeele_digital_2021,
	title = {Digital {Wellbeing} as a {Dynamic} {Construct}},
	volume = {31},
	issn = {1050-3293},
	url = {https://doi.org/10.1093/ct/qtaa024},
	doi = {10.1093/ct/qtaa024},
	number = {4},
	urldate = {2024-03-23},
	journal = {Communication Theory},
	author = {Vanden Abeele, Mariek M P},
	month = nov,
	year = {2021},
	pages = {932--955},
}

@incollection{gardner_modelling_2018,
	address = {Cham},
	title = {Modelling {Habit} {Formation} and {Its} {Determinants}},
	isbn = {978-3-319-97529-0},
	url = {https://doi.org/10.1007/978-3-319-97529-0_12},
	language = {en},
	urldate = {2024-03-15},
	booktitle = {The {Psychology} of {Habit}: {Theory}, {Mechanisms}, {Change}, and {Contexts}},
	publisher = {Springer International Publishing},
	author = {Gardner, Benjamin and Lally, Phillippa},
	editor = {Verplanken, Bas},
	year = {2018},
	doi = {10.1007/978-3-319-97529-0_12},
	pages = {207--229},
}

@article{gardner_review_2015,
	title = {A review and analysis of the use of ‘habit’ in understanding, predicting and influencing health-related behaviour},
	volume = {9},
	issn = {1743-7199},
	url = {https://www.ncbi.nlm.nih.gov/pmc/articles/PMC4566897/},
	doi = {10.1080/17437199.2013.876238},
	number = {3},
	urldate = {2024-03-15},
	journal = {Health Psychology Review},
	author = {Gardner, Benjamin},
	month = aug,
	year = {2015},
	pmid = {25207647},
	pmcid = {PMC4566897},
	pages = {277--295},
}

@article{lally_how_2010,
	title = {How are habits formed: {Modelling} habit formation in the real world},
	volume = {40},
	issn = {0046-2772, 1099-0992},
	shorttitle = {How are habits formed},
	url = {https://onlinelibrary.wiley.com/doi/10.1002/ejsp.674},
	doi = {10.1002/ejsp.674},
	language = {en},
	number = {6},
	urldate = {2024-03-15},
	journal = {European Journal of Social Psychology},
	author = {Lally, Phillippa and Van Jaarsveld, Cornelia H. M. and Potts, Henry W. W. and Wardle, Jane},
	month = oct,
	year = {2010},
	pages = {998--1009},
}

@article{wood_habit_2017,
	title = {Habit in {Personality} and {Social} {Psychology}},
	volume = {21},
	issn = {1088-8683},
	url = {https://doi.org/10.1177/1088868317720362},
	doi = {10.1177/1088868317720362},
	language = {en},
	number = {4},
	urldate = {2024-03-15},
	journal = {Personality and Social Psychology Review},
	author = {Wood, Wendy},
	month = nov,
	year = {2017},
	note = {Publisher: SAGE Publications Inc},
	pages = {389--403},
}

@article{lazer_computational_2020,
	title = {Computational social science: {Obstacles} and opportunities},
	volume = {369},
	shorttitle = {Computational social science},
	url = {https://www.science.org/doi/full/10.1126/science.aaz8170},
	doi = {10.1126/science.aaz8170},
	number = {6507},
	urldate = {2024-02-03},
	journal = {Science},
	author = {Lazer, David M. J. and Pentland, Alex and Watts, Duncan J. and Aral, Sinan and Athey, Susan and Contractor, Noshir and Freelon, Deen and Gonzalez-Bailon, Sandra and King, Gary and Margetts, Helen and Nelson, Alondra and Salganik, Matthew J. and Strohmaier, Markus and Vespignani, Alessandro and Wagner, Claudia},
	month = aug,
	year = {2020},
	note = {Publisher: American Association for the Advancement of Science},
	pages = {1060--1062},
}

@article{xu_predicting_2020,
	title = {Predicting and {Recommending} the next {Smartphone} {Apps} based on {Recurrent} {Neural} {Network}},
	volume = {2},
	issn = {2524-5228},
	url = {https://doi.org/10.1007/s42486-020-00045-z},
	doi = {10.1007/s42486-020-00045-z},
	language = {en},
	number = {4},
	urldate = {2024-01-31},
	journal = {CCF Transactions on Pervasive Computing and Interaction},
	author = {Xu, Shijian and Li, Wenzhong and Zhang, Xiao and Gao, Songcheng and Zhan, Tong and Lu, Sanglu},
	month = dec,
	year = {2020},
	pages = {314--328},
}

@article{molenaar_manifesto_2004,
	title = {A {Manifesto} on {Psychology} as {Idiographic} {Science}: {Bringing} the {Person} {Back} {Into} {Scientific} {Psychology}, {This} {Time} {Forever}},
	volume = {2},
	shorttitle = {A {Manifesto} on {Psychology} as {Idiographic} {Science}},
	doi = {10.1207/s15366359mea0204_1},
	journal = {Measurement: Interdisciplinary Research \& Perspective},
	author = {Molenaar, Peter},
	month = oct,
	year = {2004},
	pages = {201--218},
}

@article{molenaar_new_2009,
	title = {The {New} {Person}-{Specific} {Paradigm} in {Psychology}},
	volume = {18},
	issn = {0963-7214},
	url = {https://www.jstor.org/stable/20696008},
	number = {2},
	urldate = {2024-01-30},
	journal = {Current Directions in Psychological Science},
	author = {Molenaar, Peter C.M. and Campbell, Cynthia G.},
	year = {2009},
	note = {Publisher: [Association for Psychological Science, Sage Publications, Inc.]},
	pages = {112--117},
}

@article{mischel_cognitive-affective_1995,
	title = {A cognitive-affective system theory of personality: {Reconceptualizing} situations, dispositions, dynamics, and invariance in personality structure},
	volume = {102},
	issn = {1939-1471},
	shorttitle = {A cognitive-affective system theory of personality},
	doi = {10.1037/0033-295X.102.2.246},
	number = {2},
	journal = {Psychological Review},
	author = {Mischel, Walter and Shoda, Yuichi},
	year = {1995},
	note = {Place: US
Publisher: American Psychological Association},
	pages = {246--268},
}

@article{wood_psychology_2016,
	title = {Psychology of {Habit}},
	volume = {67},
	url = {https://doi.org/10.1146/annurev-psych-122414-033417},
	doi = {10.1146/annurev-psych-122414-033417},
	number = {1},
	urldate = {2024-01-29},
	journal = {Annual Review of Psychology},
	author = {Wood, Wendy and Rünger, Dennis},
	year = {2016},
	pmid = {26361052},
	note = {\_eprint: https://doi.org/10.1146/annurev-psych-122414-033417},
	pages = {289--314},
}

@article{verplanken_beyond_2006,
	title = {Beyond frequency: {Habit} as mental construct},
	volume = {45},
	copyright = {2006 The British Psychological Society},
	issn = {2044-8309},
	shorttitle = {Beyond frequency},
	url = {https://onlinelibrary.wiley.com/doi/abs/10.1348/014466605X49122},
	doi = {10.1348/014466605X49122},
	language = {en},
	number = {3},
	urldate = {2024-01-29},
	journal = {British Journal of Social Psychology},
	author = {Verplanken, Bas},
	year = {2006},
	note = {\_eprint: https://onlinelibrary.wiley.com/doi/pdf/10.1348/014466605X49122},
	pages = {639--656},
}

@misc{omalley_keras_2019,
	title = {Keras {Tuner}},
	copyright = {Apache-2.0},
	url = {https://github.com/keras-team/keras-tuner},
	urldate = {2024-01-29},
	publisher = {Keras},
	author = {O'Malley, Tom and Bursztein, Elie and Long, James and Francois, Chollet and Jin, Haifeng and Invernizzi, Luca},
	year = {2019},
	note = {original-date: 2019-06-06T22:38:21Z},
}

@article{savcisens_using_2024,
	title = {Using sequences of life-events to predict human lives},
	volume = {4},
	copyright = {2023 The Author(s), under exclusive licence to Springer Nature America, Inc.},
	issn = {2662-8457},
	url = {https://www.nature.com/articles/s43588-023-00573-5},
	doi = {10.1038/s43588-023-00573-5},
	language = {en},
	number = {1},
	urldate = {2024-01-29},
	journal = {Nature Computational Science},
	author = {Savcisens, Germans and Eliassi-Rad, Tina and Hansen, Lars Kai and Mortensen, Laust Hvas and Lilleholt, Lau and Rogers, Anna and Zettler, Ingo and Lehmann, Sune},
	month = jan,
	year = {2024},
	note = {Number: 1
Publisher: Nature Publishing Group},
	pages = {43--56},
}

@article{li_behrt_2020,
	title = {{BEHRT}: {Transformer} for {Electronic} {Health} {Records}},
	volume = {10},
	copyright = {2020 The Author(s)},
	issn = {2045-2322},
	shorttitle = {{BEHRT}},
	url = {https://www.nature.com/articles/s41598-020-62922-y},
	doi = {10.1038/s41598-020-62922-y},
	language = {en},
	number = {1},
	urldate = {2024-01-29},
	journal = {Scientific Reports},
	author = {Li, Yikuan and Rao, Shishir and Solares, José Roberto Ayala and Hassaine, Abdelaali and Ramakrishnan, Rema and Canoy, Dexter and Zhu, Yajie and Rahimi, Kazem and Salimi-Khorshidi, Gholamreza},
	month = apr,
	year = {2020},
	note = {Number: 1
Publisher: Nature Publishing Group},
	pages = {7155},
}

@article{beck_personalized_2022,
	title = {Personalized {Prediction} of {Behaviors} and {Experiences}: {An} {Idiographic} {Person}–{Situation} {Test}},
	volume = {33},
	issn = {0956-7976},
	shorttitle = {Personalized {Prediction} of {Behaviors} and {Experiences}},
	url = {https://doi.org/10.1177/09567976221093307},
	doi = {10.1177/09567976221093307},
	language = {en},
	number = {10},
	urldate = {2024-01-29},
	journal = {Psychological Science},
	author = {Beck, Emorie D. and Jackson, Joshua J.},
	month = oct,
	year = {2022},
	note = {Publisher: SAGE Publications Inc},
	pages = {1767--1782},
}

@article{bayer_consciousness_2016,
	title = {Consciousness and {Self}-{Regulation} in {Mobile} {Communication}},
	volume = {42},
	issn = {0360-3989},
	url = {https://doi.org/10.1111/hcre.12067},
	doi = {10.1111/hcre.12067},
	number = {1},
	urldate = {2024-01-29},
	journal = {Human Communication Research},
	author = {Bayer, Joseph B. and Dal Cin, Sonya and Campbell, Scott W. and Panek, Elliot},
	month = jan,
	year = {2016},
	pages = {71--97},
}

@article{verduyn_passive_2015,
	title = {Passive {Facebook} usage undermines affective well-being: {Experimental} and longitudinal evidence},
	volume = {144},
	issn = {1939-2222},
	shorttitle = {Passive {Facebook} usage undermines affective well-being},
	doi = {10.1037/xge0000057},
	number = {2},
	journal = {Journal of Experimental Psychology: General},
	author = {Verduyn, Philippe and Lee, David Seungjae and Park, Jiyoung and Shablack, Holly and Orvell, Ariana and Bayer, Joseph and Ybarra, Oscar and Jonides, John and Kross, Ethan},
	year = {2015},
	note = {Place: US
Publisher: American Psychological Association},
	pages = {480--488},
}

@article{fox_dark_2015,
	title = {The dark side of social networking sites: {An} exploration of the relational and psychological stressors associated with {Facebook} use and affordances},
	volume = {45},
	issn = {0747-5632},
	shorttitle = {The dark side of social networking sites},
	url = {https://www.sciencedirect.com/science/article/pii/S0747563214007018},
	doi = {10.1016/j.chb.2014.11.083},
	urldate = {2024-01-29},
	journal = {Computers in Human Behavior},
	author = {Fox, Jesse and Moreland, Jennifer J.},
	month = apr,
	year = {2015},
	pages = {168--176},
}

@article{primack_social_2017,
	title = {Social {Media} {Use} and {Perceived} {Social} {Isolation} {Among} {Young} {Adults} in the {U}.{S}.},
	volume = {53},
	issn = {0749-3797},
	url = {https://www.sciencedirect.com/science/article/pii/S0749379717300168},
	doi = {10.1016/j.amepre.2017.01.010},
	number = {1},
	urldate = {2024-01-29},
	journal = {American Journal of Preventive Medicine},
	author = {Primack, Brian A. and Shensa, Ariel and Sidani, Jaime E. and Whaite, Erin O. and Lin, Liu yi and Rosen, Daniel and Colditz, Jason B. and Radovic, Ana and Miller, Elizabeth},
	month = jul,
	year = {2017},
	pages = {1--8},
}

@article{allcott_welfare_2020,
	title = {The {Welfare} {Effects} of {Social} {Media}},
	volume = {110},
	issn = {0002-8282},
	url = {https://www.aeaweb.org/articles?id=10.1257%2Faer.20190658&utm_campaign=Johannes},
	doi = {10.1257/aer.20190658},
	language = {en},
	number = {3},
	urldate = {2024-01-29},
	journal = {American Economic Review},
	author = {Allcott, Hunt and Braghieri, Luca and Eichmeyer, Sarah and Gentzkow, Matthew},
	month = mar,
	year = {2020},
	pages = {629--676},
}

@article{hunt_no_2018,
	title = {No {More} {FOMO}: {Limiting} {Social} {Media} {Decreases} {Loneliness} and {Depression}},
	volume = {37},
	issn = {0736-7236},
	shorttitle = {No {More} {FOMO}},
	url = {https://guilfordjournals.com/doi/10.1521/jscp.2018.37.10.751},
	doi = {10.1521/jscp.2018.37.10.751},
	number = {10},
	urldate = {2024-01-29},
	journal = {Journal of Social and Clinical Psychology},
	author = {Hunt, Melissa G. and Marx, Rachel and Lipson, Courtney and Young, Jordyn},
	month = dec,
	year = {2018},
	note = {Publisher: Guilford Publications Inc.},
	pages = {751--768},
}

@article{brailovskaia_less_2020,
	title = {Less {Facebook} use – {More} well-being and a healthier lifestyle? {An} experimental intervention study},
	volume = {108},
	issn = {0747-5632},
	shorttitle = {Less {Facebook} use – {More} well-being and a healthier lifestyle?},
	url = {https://www.sciencedirect.com/science/article/pii/S0747563220300868},
	doi = {10.1016/j.chb.2020.106332},
	urldate = {2024-01-29},
	journal = {Computers in Human Behavior},
	author = {Brailovskaia, Julia and Ströse, Fabienne and Schillack, Holger and Margraf, Jürgen},
	month = jul,
	year = {2020},
	pages = {106332},
}

@article{licoppe_are_2005,
	series = {The {Dynamics} of {Personal} {Networks}},
	title = {Are social networks technologically embedded?: {How} networks are changing today with changes in communication technology},
	volume = {27},
	issn = {0378-8733},
	shorttitle = {Are social networks technologically embedded?},
	url = {https://www.sciencedirect.com/science/article/pii/S0378873304000619},
	doi = {10.1016/j.socnet.2004.11.001},
	number = {4},
	urldate = {2024-01-29},
	journal = {Social Networks},
	author = {Licoppe, Christian and Smoreda, Zbigniew},
	month = oct,
	year = {2005},
	pages = {317--335},
}

@article{coyle_social_2008,
	title = {Social networking: {Communication} revolution or evolution?},
	volume = {13},
	issn = {1538-7305},
	shorttitle = {Social networking},
	url = {https://ieeexplore.ieee.org/abstract/document/6771446?casa_token=Vv1NKojeijcAAAAA:mGf1Hlo0TlYHXyD3NLQxx3aYoiYyLLtY-bUuXHtNSpcGMVWu2UG3LnO8B8eeCD2XWs7oAWVATA},
	doi = {10.1002/bltj.20298},
	number = {2},
	urldate = {2024-01-29},
	journal = {Bell Labs Technical Journal},
	author = {Coyle, Cheryl L. and Vaughn, Heather},
	year = {2008},
	note = {Conference Name: Bell Labs Technical Journal},
	pages = {13--17},
}

@article{zeitzoff_how_2017,
	title = {How {Social} {Media} {Is} {Changing} {Conflict}},
	volume = {61},
	issn = {0022-0027},
	url = {https://doi.org/10.1177/0022002717721392},
	doi = {10.1177/0022002717721392},
	language = {en},
	number = {9},
	urldate = {2024-01-29},
	journal = {Journal of Conflict Resolution},
	author = {Zeitzoff, Thomas},
	month = oct,
	year = {2017},
	note = {Publisher: SAGE Publications Inc},
	pages = {1970--1991},
}

@article{diehl_political_2016,
	title = {Political persuasion on social media: {Tracing} direct and indirect effects of news use and social interaction},
	volume = {18},
	issn = {1461-4448},
	shorttitle = {Political persuasion on social media},
	url = {https://doi.org/10.1177/1461444815616224},
	doi = {10.1177/1461444815616224},
	language = {en},
	number = {9},
	urldate = {2024-01-29},
	journal = {New Media \& Society},
	author = {Diehl, Trevor and Weeks, Brian E and Gil de Zúñiga, Homero},
	month = oct,
	year = {2016},
	note = {Publisher: SAGE Publications},
	pages = {1875--1895},
}

@misc{walker_news_2021,
	title = {News {Consumption} {Across} {Social} {Media} in 2021},
	url = {https://www.pewresearch.org/journalism/2021/09/20/news-consumption-across-social-media-in-2021/},
	language = {en-US},
	urldate = {2024-01-29},
	journal = {Pew Research Center's Journalism Project},
	author = {Walker, Mason and Matsa, Katerina Eva},
	month = sep,
	year = {2021},
}

@article{gil_de_zuniga_effects_2017,
	title = {Effects of the {News}-{Finds}-{Me} {Perception} in {Communication}: {Social} {Media} {Use} {Implications} for {News} {Seeking} and {Learning} {About} {Politics}},
	volume = {22},
	issn = {1083-6101},
	shorttitle = {Effects of the {News}-{Finds}-{Me} {Perception} in {Communication}},
	url = {https://doi.org/10.1111/jcc4.12185},
	doi = {10.1111/jcc4.12185},
	number = {3},
	urldate = {2024-01-29},
	journal = {Journal of Computer-Mediated Communication},
	author = {Gil de Zúñiga, Homero and Weeks, Brian and Ardèvol-Abreu, Alberto},
	month = may,
	year = {2017},
	pages = {105--123},
}

@article{vermeer_online_2020,
	title = {Online {News} {User} {Journeys}: {The} {Role} of {Social} {Media}, {News} {Websites}, and {Topics}},
	volume = {8},
	issn = {2167-0811},
	shorttitle = {Online {News} {User} {Journeys}},
	url = {https://doi.org/10.1080/21670811.2020.1767509},
	doi = {10.1080/21670811.2020.1767509},
	number = {9},
	urldate = {2024-01-29},
	journal = {Digital Journalism},
	author = {Vermeer, Susan and Trilling, Damian and Kruikemeier, Sanne and de Vreese, Claes},
	month = oct,
	year = {2020},
	note = {Publisher: Routledge
\_eprint: https://doi.org/10.1080/21670811.2020.1767509},
	pages = {1114--1141},
}

@article{lee_effects_2017,
	title = {The effects of news consumption via social media and news information overload on perceptions of journalistic norms and practices},
	volume = {75},
	issn = {0747-5632},
	url = {https://www.sciencedirect.com/science/article/pii/S0747563217303199},
	doi = {10.1016/j.chb.2017.05.007},
	urldate = {2024-01-29},
	journal = {Computers in Human Behavior},
	author = {Lee, Sun Kyong and Lindsey, Nathan J. and Kim, Kyun Soo},
	month = oct,
	year = {2017},
	pages = {254--263},
}

@article{gruning_directing_2023,
	title = {Directing smartphone use through the self-nudge app one sec},
	volume = {120},
	url = {https://www.pnas.org/doi/10.1073/pnas.2213114120},
	doi = {10.1073/pnas.2213114120},
	number = {8},
	urldate = {2024-01-28},
	journal = {Proceedings of the National Academy of Sciences},
	author = {Grüning, David J. and Riedel, Frederik and Lorenz-Spreen, Philipp},
	month = feb,
	year = {2023},
	note = {Publisher: Proceedings of the National Academy of Sciences},
	pages = {e2213114120},
}

@misc{statista_us_2024,
	title = {U.{S}.: mobile phone daily usage time 2024},
	shorttitle = {U.{S}.},
	url = {https://www.statista.com/statistics/1045353/mobile-device-daily-usage-time-in-the-us/},
	language = {en},
	urldate = {2024-01-26},
	journal = {Statista},
	author = {Statista},
	year = {2024},
}

@misc{openai_gpt-4_2023,
	title = {{GPT}-4 {Technical} {Report}},
	url = {https://cdn.openai.com/papers/gpt-4.pdf},
	urldate = {2023-08-21},
	author = {OpenAI},
	year = {2023},
}

@article{ward_brain_2017,
	title = {Brain {Drain}: {The} {Mere} {Presence} of {One}’s {Own} {Smartphone} {Reduces} {Available} {Cognitive} {Capacity}},
	volume = {2},
	issn = {2378-1815},
	shorttitle = {Brain {Drain}},
	url = {https://www.journals.uchicago.edu/doi/full/10.1086/691462},
	doi = {10.1086/691462},
	number = {2},
	urldate = {2024-01-24},
	journal = {Journal of the Association for Consumer Research},
	author = {Ward, Adrian F. and Duke, Kristen and Gneezy, Ayelet and Bos, Maarten W.},
	month = apr,
	year = {2017},
	note = {Publisher: The University of Chicago Press},
	pages = {140--154},
}

@article{duke_smartphone_2017,
	title = {Smartphone addiction, daily interruptions and self-reported productivity},
	volume = {6},
	issn = {2352-8532},
	url = {https://www.ncbi.nlm.nih.gov/pmc/articles/PMC5800562/},
	doi = {10.1016/j.abrep.2017.07.002},
	urldate = {2024-01-24},
	journal = {Addictive Behaviors Reports},
	author = {Duke, Éilish and Montag, Christian},
	month = jul,
	year = {2017},
	pmid = {29450241},
	pmcid = {PMC5800562},
	pages = {90--95},
}

@article{derks_private_2021,
	title = {Private smartphone use during worktime: {A} diary study on the unexplored costs of integrating the work and family domains},
	volume = {114},
	issn = {0747-5632},
	shorttitle = {Private smartphone use during worktime},
	url = {https://www.sciencedirect.com/science/article/pii/S074756322030282X},
	doi = {10.1016/j.chb.2020.106530},
	urldate = {2024-01-24},
	journal = {Computers in Human Behavior},
	author = {Derks, Daantje and Bakker, Arnold B. and Gorgievski, Marjan},
	month = jan,
	year = {2021},
	pages = {106530},
}

@article{reeves_screenomics_2021,
	title = {Screenomics: {A} {Framework} to {Capture} and {Analyze} {Personal} {Life} {Experiences} and the {Ways} that {Technology} {Shapes} {Them}},
	volume = {36},
	issn = {0737-0024},
	shorttitle = {Screenomics},
	url = {https://doi.org/10.1080/07370024.2019.1578652},
	doi = {10.1080/07370024.2019.1578652},
	number = {2},
	urldate = {2023-12-14},
	journal = {Human–Computer Interaction},
	author = {Reeves, Byron and Ram, Nilam and Robinson, Thomas N. and Cummings, James J. and Giles, C. Lee and Pan, Jennifer and Chiatti, Agnese and Cho, Mj and Roehrick, Katie and Yang, Xiao and Gagneja, Anupriya and Brinberg, Miriam and Muise, Daniel and Lu, Yingdan and Luo, Mufan and Fitzgerald, Andrew and Yeykelis, Leo},
	month = mar,
	year = {2021},
	pmid = {33867652},
	note = {Publisher: Taylor \& Francis
\_eprint: https://doi.org/10.1080/07370024.2019.1578652},
	pages = {150--201},
}

@article{boyle_systematic_2022,
	title = {Systematic {Bias} in {Self}-{Reported} {Social} {Media} {Use} in the {Age} of {Platform} {Swinging}: {Implications} for {Studying} {Social} {Media} {Use} in {Relation} to {Adolescent} {Health} {Behavior}},
	volume = {19},
	issn = {1661-7827},
	shorttitle = {Systematic {Bias} in {Self}-{Reported} {Social} {Media} {Use} in the {Age} of {Platform} {Swinging}},
	url = {https://www.ncbi.nlm.nih.gov/pmc/articles/PMC9408042/},
	doi = {10.3390/ijerph19169847},
	number = {16},
	urldate = {2023-12-14},
	journal = {International Journal of Environmental Research and Public Health},
	author = {Boyle, Sarah C. and Baez, Sebastian and Trager, Bradley M. and LaBrie, Joseph W.},
	month = aug,
	year = {2022},
	pmid = {36011479},
	pmcid = {PMC9408042},
	pages = {9847},
}

@article{haslbeck_recovering_2022,
	title = {Recovering {Within}-{Person} {Dynamics} from {Psychological} {Time} {Series}},
	volume = {57},
	issn = {0027-3171},
	url = {https://doi.org/10.1080/00273171.2021.1896353},
	doi = {10.1080/00273171.2021.1896353},
	number = {5},
	urldate = {2023-12-14},
	journal = {Multivariate Behavioral Research},
	author = {Haslbeck, Jonas M. B. and Ryan, Oisín},
	month = sep,
	year = {2022},
	pmid = {34154483},
	note = {Publisher: Routledge
\_eprint: https://doi.org/10.1080/00273171.2021.1896353},
	pages = {735--766},
}

@article{verplanken_reflections_2003,
	title = {Reflections on {Past} {Behavior}: {A} {Self}-{Report} {Index} of {Habit} {Strength}},
	volume = {33},
	issn = {1559-1816},
	shorttitle = {Reflections on {Past} {Behavior}},
	url = {https://onlinelibrary.wiley.com/doi/abs/10.1111/j.1559-1816.2003.tb01951.x},
	doi = {10.1111/j.1559-1816.2003.tb01951.x},
	language = {en},
	number = {6},
	urldate = {2023-11-10},
	journal = {Journal of Applied Social Psychology},
	author = {Verplanken, Bas and Orbell, Sheina},
	year = {2003},
	note = {\_eprint: https://onlinelibrary.wiley.com/doi/pdf/10.1111/j.1559-1816.2003.tb01951.x},
	pages = {1313--1330},
}

@article{gardner_towards_2012,
	title = {Towards parsimony in habit measurement: {Testing} the convergent and predictive validity of an automaticity subscale of the {Self}-{Report} {Habit} {Index}},
	volume = {9},
	issn = {1479-5868},
	shorttitle = {Towards parsimony in habit measurement},
	url = {https://doi.org/10.1186/1479-5868-9-102},
	doi = {10.1186/1479-5868-9-102},
	number = {1},
	urldate = {2023-11-10},
	journal = {International Journal of Behavioral Nutrition and Physical Activity},
	author = {Gardner, Benjamin and Abraham, Charles and Lally, Phillippa and de Bruijn, Gert-Jan},
	month = aug,
	year = {2012},
	pages = {102},
}

@misc{pedregosa_scikit-learn_2018,
	title = {Scikit-learn: {Machine} {Learning} in {Python}},
	shorttitle = {Scikit-learn},
	url = {http://arxiv.org/abs/1201.0490},
	doi = {10.48550/arXiv.1201.0490},
	urldate = {2023-02-21},
	publisher = {arXiv},
	author = {Pedregosa, Fabian and Varoquaux, Gaël and Gramfort, Alexandre and Michel, Vincent and Thirion, Bertrand and Grisel, Olivier and Blondel, Mathieu and Müller, Andreas and Nothman, Joel and Louppe, Gilles and Prettenhofer, Peter and Weiss, Ron and Dubourg, Vincent and Vanderplas, Jake and Passos, Alexandre and Cournapeau, David and Brucher, Matthieu and Perrot, Matthieu and Duchesnay, Édouard},
	month = jun,
	year = {2018},
	note = {arXiv:1201.0490 [cs]},
}

@article{tibshirani_regression_1996,
	title = {Regression {Shrinkage} and {Selection} via the {Lasso}},
	volume = {58},
	issn = {0035-9246},
	url = {https://www.jstor.org/stable/2346178},
	number = {1},
	urldate = {2023-01-30},
	journal = {Journal of the Royal Statistical Society. Series B (Methodological)},
	author = {Tibshirani, Robert},
	year = {1996},
	note = {Publisher: [Royal Statistical Society, Wiley]},
	pages = {267--288},
}

@article{breiman_random_2001,
	title = {Random {Forests}},
	volume = {45},
	issn = {1573-0565},
	url = {https://doi.org/10.1023/A:1010933404324},
	doi = {10.1023/A:1010933404324},
	language = {en},
	number = {1},
	urldate = {2023-01-30},
	journal = {Machine Learning},
	author = {Breiman, Leo},
	month = oct,
	year = {2001},
	pages = {5--32},
}

@article{harari_using_2016,
	title = {Using {Smartphones} to {Collect} {Behavioral} {Data} in {Psychological} {Science}: {Opportunities}, {Practical} {Considerations}, and {Challenges}},
	volume = {11},
	issn = {1745-6916},
	shorttitle = {Using {Smartphones} to {Collect} {Behavioral} {Data} in {Psychological} {Science}},
	url = {https://doi.org/10.1177/1745691616650285},
	doi = {10.1177/1745691616650285},
	language = {en},
	number = {6},
	urldate = {2022-04-14},
	journal = {Perspectives on Psychological Science},
	author = {Harari, Gabriella M. and Lane, Nicholas D. and Wang, Rui and Crosier, Benjamin S. and Campbell, Andrew T. and Gosling, Samuel D.},
	month = nov,
	year = {2016},
	note = {Publisher: SAGE Publications Inc},
	pages = {838--854},
}

@article{stachl_personality_2020,
	title = {Personality {Research} and {Assessment} in the {Era} of {Machine} {Learning}},
	volume = {34},
	copyright = {© 2020 The Authors. European Journal of Personality published by John Wiley \& Sons Ltd on behalf of European Association of Personality Psychology},
	issn = {1099-0984},
	url = {https://onlinelibrary.wiley.com/doi/abs/10.1002/per.2257},
	doi = {https://doi.org/10.1002/per.2257},
	language = {en},
	number = {5},
	urldate = {2020-12-01},
	journal = {European Journal of Personality},
	author = {Stachl, Clemens and Pargent, Florian and Hilbert, Sven and Harari, Gabriella M. and Schoedel, Ramona and Vaid, Sumer and Gosling, Samuel D. and Bühner, Markus},
	year = {2020},
	note = {\_eprint: https://onlinelibrary.wiley.com/doi/pdf/10.1002/per.2257},
	pages = {613--631},
}

@article{mischel_toward_1974,
	title = {Toward a cognitive social learning reconceptualization of personality.},
	volume = {80},
	issn = {1939-1471},
	url = {http://psycnet.apa.org/fulltext/1974-06274-001.pdf},
	doi = {10.1037/h0035002},
	number = {4},
	urldate = {2019-11-28},
	journal = {Psychological Review},
	author = {Mischel, Walter},
	year = {1974},
	pages = {252},
}

\end{document}


\maketitle


\newpage
\appendix

\section{List of Social Media Apps}
\label{habits_social_media_apps}
\begin{itemize}
    \item Discord
    \item Facebook
    \item Facebook Lite
    \item Facebook Local
    \item Instagram
    \item Kik
    \item LinkedIn
    \item Pinterest
    \item Reddit
    \item Snapchat
    \item TikTok
    \item Tumblr
    \item Twitter
    \item Youtube
\end{itemize}

\FloatBarrier
\newpage

\section{Hyperparameter Spaces}
\label{habits_hyperparameters}
\subsection{LSTM Models}

\begin{table}[h]
\renewcommand{\arraystretch}{1} 
\centering
\label{table:lstm-models}
\noindent\begin{tabularx}{\textwidth}{Xrrr}
\toprule
Hyperparameter Name & Min & Max & Step Size\\
\midrule
Embedding Dimensions & 5 & 50 & 5 \\
Number of LSTM Layers & 1 & 3 & 1 \\
LSTM Units & 4 & 64 & 4 \\
Recurrent Dropout & 0.2 & 0.5 & 0.1 \\
LSTM L1 Regularization & $1e-5$ & $1e-3$ & Continuous (log) \\
LSTM L2 Regularization & $1e-4$ & $1e-2$ & Continuous (log) \\
Dense Layer Units & 4 & 64 & 4 \\
Dense Layer L1 Regularization & $1e-5$ & $1e-3$ & Continuous (log) \\
Dense Layer L2 Regularization & $1e-4$ & $1e-2$ & Continuous (log) \\
Dropout Rate Top Layer & 0.2 & 0.5 & 0.1 \\
Learning Rate & $1e-5$ & $1e-2$ & Continuous (log) \\
\bottomrule
\end{tabularx}
\end{table}

\FloatBarrier

\subsection{Transformer Models}

\begin{table}[h]
\renewcommand{\arraystretch}{1} 
\centering
\label{table:lstm-models}
\noindent\begin{tabularx}{\textwidth}{Xrrr}
\toprule
Hyperparameter Name & Min & Max & Step Size\\
\midrule
Embedding Dimensions & 5 & 50 & 5 \\
Number of Transformer Layers* & 1 & 3 & 1 \\
Transformer Units & 4 & 64 & 4 \\
Recurrent Dropout*& 0.2 & 0.5 & 0.1 \\
Transformer L1 Regularization & $1e-5$ & $1e-3$ & Continuous (log) \\
Transformer L2 Regularization & $1e-4$ & $1e-2$ & Continuous (log) \\
Dense Layer Units & 4 & 64 & 4 \\
Dense Layer L1 Regularization & $1e-5$ & $1e-3$ & Continuous (log) \\
Dense Layer L2 Regularization & $1e-4$ & $1e-2$ & Continuous (log) \\
Dropout Rate Top Layer & 0.2 & 0.5 & 0.1 \\
Learning Rate & $1e-4$ & $1e-2$ & Continuous (log) \\
\bottomrule
\end{tabularx}
\end{table}


\subsection{Fine-Tuned Models}

\begin{table}[h]
\renewcommand{\arraystretch}{1} 
\centering
\label{table:lstm-models}
\noindent\begin{tabularx}{\textwidth}{Xrrr}
\toprule
Hyperparameter Name & Min & Max & Step Size\\
\midrule
Dense Layer Units & 4 & 64 & 4 \\
Dense Layer L1 Regularization & $1e-5$ & $1e-3$ & Continuous (log) \\
Dense Layer L2 Regularization & $1e-4$ & $1e-2$ & Continuous (log) \\
Dropout Rate Top Layer & 0.2 & 0.5 & 0.1 \\
Learning Rate & $1e-4$ & $1e-2$ & Continuous (log) \\
\bottomrule
\end{tabularx}
\end{table}

\FloatBarrier
\newpage

\section{Additional Model Evaluation Metrics for Main Analyses}
\label{habits_all_metrics}
\subsection{Performance of Global LSTM Model Across 20 Rounds of Hyperparameter Search}

\begin{table}[h]
\begin{tabular*}{\textwidth}{@{\extracolsep{\fill}}lrrrrrr}
\toprule
rank & acc & pre & rec & f1 & auc \\
\midrule
1 & 0.779 & 0.713 & 0.651 & 0.668 & 0.782 \\
2 & 0.781 & 0.728 & 0.631 & 0.648 & 0.780 \\
3 & 0.780 & 0.716 & 0.652 & 0.669 & 0.778 \\
4 & 0.769 & 0.700 & 0.615 & 0.628 & 0.764 \\
5 & 0.771 & 0.705 & 0.618 & 0.632 & 0.763 \\
6 & 0.765 & 0.728 & 0.571 & 0.566 & 0.761 \\
7 & 0.770 & 0.707 & 0.607 & 0.618 & 0.761 \\
8 & 0.768 & 0.705 & 0.605 & 0.615 & 0.760 \\
9 & 0.769 & 0.706 & 0.605 & 0.616 & 0.759 \\
10 & 0.767 & 0.695 & 0.617 & 0.629 & 0.758 \\
11 & 0.768 & 0.695 & 0.625 & 0.639 & 0.759 \\
12 & 0.769 & 0.705 & 0.604 & 0.615 & 0.758 \\
13 & 0.769 & 0.701 & 0.613 & 0.626 & 0.757 \\
14 & 0.768 & 0.703 & 0.604 & 0.614 & 0.755 \\
15 & 0.768 & 0.702 & 0.608 & 0.620 & 0.754 \\
16 & 0.769 & 0.704 & 0.608 & 0.620 & 0.753 \\
17 & 0.767 & 0.702 & 0.598 & 0.607 & 0.749 \\
18 & 0.741 & 0.371 & 0.500 & 0.426 & 0.507 \\
19 & 0.741 & 0.371 & 0.500 & 0.426 & 0.500 \\
20 & 0.741 & 0.371 & 0.500 & 0.426 & 0.568 \\
\bottomrule
\end{tabular*}
\end{table}

\FloatBarrier

\subsection{Performance of Global Transformer Model Across 20 Rounds of Hyperparameter Search}

\begin{table}[h]
\begin{tabular*}{\textwidth}{@{\extracolsep{\fill}}lrrrrrr}
\toprule
rank & acc & pre & rec & f1 & auc \\
\midrule
1 & 0.775 & 0.706 & 0.643 & 0.659 & 0.773 \\
2 & 0.777 & 0.716 & 0.628 & 0.644 & 0.775 \\
3 & 0.771 & 0.697 & 0.659 & 0.672 & 0.774 \\
4 & 0.776 & 0.712 & 0.636 & 0.652 & 0.774 \\
5 & 0.774 & 0.709 & 0.630 & 0.645 & 0.773 \\
6 & 0.776 & 0.716 & 0.626 & 0.641 & 0.773 \\
7 & 0.775 & 0.707 & 0.645 & 0.661 & 0.774 \\
8 & 0.775 & 0.717 & 0.621 & 0.635 & 0.773 \\
9 & 0.775 & 0.709 & 0.636 & 0.651 & 0.771 \\
10 & 0.776 & 0.715 & 0.626 & 0.641 & 0.774 \\
11 & 0.773 & 0.701 & 0.646 & 0.661 & 0.772 \\
12 & 0.775 & 0.712 & 0.626 & 0.641 & 0.772 \\
13 & 0.773 & 0.703 & 0.642 & 0.658 & 0.772 \\
14 & 0.775 & 0.708 & 0.636 & 0.651 & 0.773 \\
15 & 0.775 & 0.710 & 0.634 & 0.649 & 0.773 \\
16 & 0.775 & 0.710 & 0.633 & 0.649 & 0.773 \\
17 & 0.772 & 0.707 & 0.620 & 0.634 & 0.767 \\
18 & 0.774 & 0.711 & 0.621 & 0.635 & 0.769 \\
19 & 0.774 & 0.716 & 0.614 & 0.627 & 0.771 \\
20 & 0.771 & 0.709 & 0.613 & 0.625 & 0.766 \\
\bottomrule
\end{tabular*}
\end{table}

\FloatBarrier
\newpage

\subsection{Distributions of Model Performance Scores (LSTM, Global Model)}

\begin{table}[h]
    \begin{tabular*}{\textwidth}{@{\extracolsep{\fill}}lrrrrr}
\toprule
 & auc & acc & pre & rec & f1 \\
\midrule
mean & 0.699 & 0.785 & 0.627 & 0.580 & 0.578 \\
std & 0.088 & 0.100 & 0.106 & 0.075 & 0.092 \\
min & 0.515 & 0.552 & 0.386 & 0.486 & 0.432 \\
25\% & 0.635 & 0.714 & 0.570 & 0.522 & 0.508 \\
50\% & 0.690 & 0.795 & 0.621 & 0.556 & 0.554 \\
75\% & 0.752 & 0.875 & 0.691 & 0.631 & 0.645 \\
max & 0.908 & 0.978 & 0.886 & 0.798 & 0.806 \\
\bottomrule
\end{tabular*}
\end{table}

\FloatBarrier

\vspace{0.5cm}
\subsection{Distributions of Model Performance Scores (Transformer, Global Model)}

\begin{table}[h]
    \begin{tabular*}{\textwidth}{@{\extracolsep{\fill}}lrrrrr}
\toprule
 & auc & acc & pre & rec & f1 \\
\midrule
mean & 0.679 & 0.779 & 0.620 & 0.560 & 0.554 \\
std & 0.072 & 0.103 & 0.089 & 0.054 & 0.068 \\
min & 0.507 & 0.566 & 0.408 & 0.494 & 0.447 \\
25\% & 0.630 & 0.701 & 0.579 & 0.515 & 0.493 \\
50\% & 0.677 & 0.768 & 0.629 & 0.546 & 0.543 \\
75\% & 0.724 & 0.855 & 0.669 & 0.592 & 0.597 \\
max & 0.853 & 0.981 & 0.823 & 0.703 & 0.728 \\
\bottomrule
\end{tabular*}
\end{table}

\FloatBarrier

\subsection{Distributions of Model Performance Scores (LSTM, Person-Specific Models)}

\begin{table}[h]
    \begin{tabular*}{\textwidth}{@{\extracolsep{\fill}}lrrrrr}
\toprule
 & auc & acc & pre & rec & f1 \\
\midrule
mean & 0.609 & 0.772 & 0.451 & 0.520 & 0.469 \\
std & 0.092 & 0.114 & 0.114 & 0.051 & 0.080 \\
min & 0.414 & 0.422 & 0.287 & 0.483 & 0.320 \\
25\% & 0.544 & 0.689 & 0.374 & 0.500 & 0.425 \\
50\% & 0.611 & 0.776 & 0.415 & 0.500 & 0.451 \\
75\% & 0.681 & 0.857 & 0.468 & 0.500 & 0.479 \\
max & 0.817 & 0.981 & 0.829 & 0.724 & 0.740 \\
\bottomrule
\end{tabular*}
\end{table}

\FloatBarrier

\subsection{Distributions of Model Performance Scores (Transformer, Person-Specific Models)}

\begin{table}[h]
    \begin{tabular*}{\textwidth}{@{\extracolsep{\fill}}lrrrrr}
\toprule
 & auc & acc & pre & rec & f1 \\
\midrule
mean & 0.627 & 0.772 & 0.537 & 0.539 & 0.511 \\
std & 0.087 & 0.105 & 0.136 & 0.059 & 0.083 \\
min & 0.447 & 0.523 & 0.295 & 0.455 & 0.371 \\
25\% & 0.560 & 0.691 & 0.432 & 0.500 & 0.452 \\
50\% & 0.635 & 0.769 & 0.501 & 0.500 & 0.479 \\
75\% & 0.685 & 0.841 & 0.643 & 0.569 & 0.570 \\
max & 0.861 & 0.981 & 0.915 & 0.752 & 0.757 \\
\bottomrule
\end{tabular*}
\end{table}


\FloatBarrier
\newpage

\subsection{Distributions of Model Performance Scores (LSTM, Fine-Tuned Models)}

\begin{table}[h]
    \begin{tabular*}{\textwidth}{@{\extracolsep{\fill}}lrrrrr}
\toprule
 & auc & acc & pre & rec & f1 \\
\midrule
mean & 0.702 & 0.791 & 0.626 & 0.577 & 0.569 \\
std & 0.089 & 0.098 & 0.136 & 0.080 & 0.102 \\
min & 0.522 & 0.605 & 0.372 & 0.492 & 0.426 \\
25\% & 0.640 & 0.720 & 0.541 & 0.503 & 0.480 \\
50\% & 0.689 & 0.796 & 0.625 & 0.556 & 0.553 \\
75\% & 0.754 & 0.877 & 0.700 & 0.627 & 0.627 \\
max & 0.908 & 0.978 & 0.972 & 0.800 & 0.807 \\
\bottomrule
\end{tabular*}
\end{table}

\FloatBarrier
\vspace{0.5cm}
\subsection{Distributions of Model Performance Scores (Transformer, Fine-Tuned Models)}

\begin{table}[h]
    \begin{tabular*}{\textwidth}{@{\extracolsep{\fill}}lrrrrr}
\toprule
 & auc & acc & pre & rec & f1 \\
\midrule
mean & 0.683 & 0.783 & 0.625 & 0.568 & 0.560 \\
std & 0.077 & 0.102 & 0.100 & 0.068 & 0.085 \\
min & 0.497 & 0.552 & 0.388 & 0.489 & 0.437 \\
25\% & 0.635 & 0.707 & 0.572 & 0.513 & 0.486 \\
50\% & 0.681 & 0.788 & 0.630 & 0.548 & 0.552 \\
75\% & 0.731 & 0.859 & 0.688 & 0.609 & 0.616 \\
max & 0.858 & 0.981 & 0.823 & 0.753 & 0.766 \\
\bottomrule
\end{tabular*}
\end{table}

The full set of results is available for download in CSV format on this project's OSF page
(\href{https://osf.io/rkswe/}{https://osf.io/rkswe/}).

\FloatBarrier
\newpage

\section{Models Trained on Data Including Launcher and SystemUI Events}
\label{habits_alternative_results}



\subsection{Performance of Global LSTM Model Across 20 Rounds of Hyperparameter Search}

\begin{table}[h]
\begin{tabular*}{\textwidth}{@{\extracolsep{\fill}}lrrrrr}
\toprule
 rank & acc & pre & rec & f1 & auc \\
\midrule
1 & 0.847 & 0.737 & 0.622 & 0.649 & 0.826 \\
2 & 0.847 & 0.737 & 0.626 & 0.653 & 0.827 \\
3 & 0.848 & 0.747 & 0.608 & 0.634 & 0.827 \\
4 & 0.847 & 0.737 & 0.620 & 0.646 & 0.826 \\
5 & 0.847 & 0.758 & 0.590 & 0.611 & 0.826 \\
6 & 0.847 & 0.735 & 0.624 & 0.651 & 0.827 \\
7 & 0.845 & 0.734 & 0.607 & 0.632 & 0.818 \\
8 & 0.842 & 0.716 & 0.652 & 0.674 & 0.823 \\
9 & 0.845 & 0.729 & 0.616 & 0.641 & 0.822 \\
10 & 0.846 & 0.734 & 0.623 & 0.649 & 0.823 \\
11 & 0.846 & 0.737 & 0.612 & 0.637 & 0.821 \\
12 & 0.846 & 0.733 & 0.614 & 0.639 & 0.819 \\
13 & 0.838 & 0.733 & 0.551 & 0.553 & 0.793 \\
14 & 0.839 & 0.728 & 0.562 & 0.570 & 0.792 \\
15 & 0.830 & 0.415 & 0.500 & 0.454 & 0.704 \\
16 & 0.830 & 0.415 & 0.500 & 0.454 & 0.701 \\
17 & 0.830 & 0.415 & 0.500 & 0.454 & 0.692 \\
18 & 0.830 & 0.415 & 0.500 & 0.454 & 0.495 \\
19 & 0.830 & 0.415 & 0.500 & 0.454 & 0.500 \\
20 & 0.830 & 0.415 & 0.500 & 0.454 & 0.501 \\
\bottomrule
\end{tabular*}
\end{table}


\subsection{Performance of Global Transformer Model Across 20 Rounds of Hyperparameter Search}

\begin{table}[h]
\begin{tabular*}{\textwidth}{@{\extracolsep{\fill}}lrrrrr}
\toprule
rank & acc & pre & rec & f1 & auc \\
\midrule
1 & 0.846 & 0.730 & 0.636 & 0.662 & 0.825 \\
2 & 0.848 & 0.752 & 0.602 & 0.627 & 0.824 \\
3 & 0.847 & 0.741 & 0.613 & 0.639 & 0.824 \\
4 & 0.847 & 0.740 & 0.613 & 0.639 & 0.823 \\
5 & 0.847 & 0.736 & 0.621 & 0.647 & 0.823 \\
6 & 0.849 & 0.757 & 0.606 & 0.632 & 0.824 \\
7 & 0.848 & 0.743 & 0.618 & 0.644 & 0.822 \\
8 & 0.848 & 0.740 & 0.618 & 0.644 & 0.822 \\
9 & 0.849 & 0.746 & 0.622 & 0.650 & 0.823 \\
10 & 0.847 & 0.744 & 0.610 & 0.636 & 0.821 \\
11 & 0.847 & 0.745 & 0.608 & 0.634 & 0.821 \\
12 & 0.846 & 0.748 & 0.593 & 0.615 & 0.820 \\
13 & 0.847 & 0.740 & 0.610 & 0.636 & 0.817 \\
14 & 0.844 & 0.728 & 0.616 & 0.641 & 0.818 \\
15 & 0.846 & 0.735 & 0.612 & 0.638 & 0.816 \\
16 & 0.844 & 0.725 & 0.616 & 0.641 & 0.813 \\
17 & 0.839 & 0.733 & 0.557 & 0.562 & 0.802 \\
18 & 0.841 & 0.726 & 0.585 & 0.603 & 0.805 \\
19 & 0.840 & 0.719 & 0.589 & 0.609 & 0.801 \\
20 & 0.837 & 0.710 & 0.564 & 0.575 & 0.787 \\
\bottomrule
\end{tabular*}
\end{table}

\FloatBarrier
\newpage

\section{Models Trained with Weighted Loss to Adjust for Class-Imbalances}
\label{balanced_class_weights}

\subsection{Performance of Global LSTM Model Across 20 Rounds of Hyperparameter Search}
\begin{table}[h]
\begin{tabular*}{\textwidth}{@{\extracolsep{\fill}}lrrrrr}
\toprule
rank & acc & pre & rec & f1 & auc \\
\midrule
 1 & 0.728 & 0.481 & 0.658 & 0.556 & 0.781 \\
2 & 0.719 & 0.471 & 0.697 & 0.562 & 0.785 \\
3 & 0.715 & 0.466 & 0.698 & 0.559 & 0.783 \\
4 & 0.714 & 0.465 & 0.699 & 0.559 & 0.783 \\
5 & 0.720 & 0.471 & 0.690 & 0.560 & 0.784 \\
6 & 0.703 & 0.451 & 0.682 & 0.543 & 0.767 \\
7 & 0.681 & 0.429 & 0.711 & 0.535 & 0.761 \\
8 & 0.677 & 0.428 & 0.733 & 0.540 & 0.766 \\
9 & 0.685 & 0.434 & 0.719 & 0.541 & 0.765 \\
10 & 0.645 & 0.404 & 0.786 & 0.534 & 0.763 \\
11 & 0.682 & 0.430 & 0.711 & 0.536 & 0.760 \\
12 & 0.683 & 0.432 & 0.716 & 0.539 & 0.762 \\
13 & 0.697 & 0.444 & 0.680 & 0.537 & 0.760 \\
14 & 0.711 & 0.457 & 0.625 & 0.528 & 0.757 \\
15 & 0.679 & 0.428 & 0.718 & 0.536 & 0.761 \\
16 & 0.695 & 0.441 & 0.668 & 0.531 & 0.755 \\
17 & 0.680 & 0.428 & 0.711 & 0.534 & 0.758 \\
18 & 0.687 & 0.434 & 0.691 & 0.533 & 0.757 \\
19 & 0.674 & 0.413 & 0.616 & 0.494 & 0.714 \\
20 & 0.741 & 0.000 & 0.000 & 0.000 & 0.500 \\
\bottomrule
\end{tabular*}
\end{table}


\subsection{Performance of Global Transformer Model Across 20 Rounds of Hyperparameter Search}

\begin{table}[h]
\begin{tabular*}{\textwidth}{@{\extracolsep{\fill}}lrrrrr}
\toprule
rank & acc & pre & rec & f1 & auc \\
\midrule
 1 & 0.702 & 0.451 & 0.707 & 0.550 & 0.775 \\
2 & 0.726 & 0.477 & 0.644 & 0.548 & 0.774 \\
3 & 0.700 & 0.449 & 0.702 & 0.548 & 0.773 \\
4 & 0.710 & 0.459 & 0.685 & 0.550 & 0.772 \\
5 & 0.675 & 0.427 & 0.756 & 0.546 & 0.773 \\
6 & 0.684 & 0.435 & 0.741 & 0.548 & 0.774 \\
7 & 0.697 & 0.445 & 0.703 & 0.545 & 0.771 \\
8 & 0.687 & 0.437 & 0.731 & 0.547 & 0.774 \\
9 & 0.691 & 0.441 & 0.727 & 0.549 & 0.773 \\
10 & 0.708 & 0.457 & 0.681 & 0.547 & 0.773 \\
11 & 0.674 & 0.426 & 0.754 & 0.545 & 0.770 \\
12 & 0.696 & 0.445 & 0.701 & 0.544 & 0.769 \\
13 & 0.667 & 0.420 & 0.756 & 0.540 & 0.769 \\
14 & 0.670 & 0.423 & 0.751 & 0.541 & 0.768 \\
15 & 0.688 & 0.437 & 0.716 & 0.543 & 0.765 \\
16 & 0.692 & 0.439 & 0.691 & 0.537 & 0.762 \\
17 & 0.680 & 0.425 & 0.673 & 0.521 & 0.745 \\
18 & 0.698 & 0.441 & 0.632 & 0.520 & 0.745 \\
19 & 0.683 & 0.431 & 0.704 & 0.535 & 0.738 \\
20 & 0.741 & 0.000 & 0.000 & 0.000 & 0.545 \\
\bottomrule
\end{tabular*}
\end{table}

\FloatBarrier
\newpage

\subsection{Distributions of Model Performance Scores (LSTM, Global Model)}

\begin{table}[h]
    \begin{tabular*}{\textwidth}{@{\extracolsep{\fill}}lrrrrr}
\toprule
 & auc & acc & pre & rec & f1 \\
\midrule
mean & 0.694 & 0.732 & 0.621 & 0.623 & 0.605 \\
std & 0.088 & 0.113 & 0.060 & 0.075 & 0.066 \\
min & 0.502 & 0.522 & 0.514 & 0.505 & 0.450 \\
25\% & 0.634 & 0.648 & 0.578 & 0.573 & 0.565 \\
50\% & 0.678 & 0.722 & 0.610 & 0.612 & 0.598 \\
75\% & 0.739 & 0.815 & 0.668 & 0.656 & 0.645 \\
max & 0.909 & 0.969 & 0.776 & 0.828 & 0.774 \\
\bottomrule
\end{tabular*}
\end{table}

\FloatBarrier

\subsection{Distributions of Model Performance Scores (Transformer, Global Model)}

\begin{table}[h]
    \begin{tabular*}{\textwidth}{@{\extracolsep{\fill}}lrrrrr}
\toprule
 & auc & acc & pre & rec & f1 \\
\midrule
mean & 0.681 & 0.705 & 0.611 & 0.612 & 0.583 \\
std & 0.076 & 0.124 & 0.058 & 0.064 & 0.064 \\
min & 0.519 & 0.479 & 0.490 & 0.492 & 0.457 \\
25\% & 0.626 & 0.605 & 0.567 & 0.561 & 0.534 \\
50\% & 0.679 & 0.682 & 0.605 & 0.616 & 0.585 \\
75\% & 0.730 & 0.803 & 0.649 & 0.649 & 0.618 \\
max & 0.848 & 0.964 & 0.810 & 0.773 & 0.724 \\
\bottomrule
\end{tabular*}
\end{table}

\FloatBarrier

\subsection{Distributions of Model Performance Scores (LSTM, Person-Specific Models)}

\begin{table}[h]
    \begin{tabular*}{\textwidth}{@{\extracolsep{\fill}}lrrrrr}
\toprule
 & auc & acc & pre & rec & f1 \\
\midrule
mean & 0.630 & 0.595 & 0.568 & 0.586 & 0.507 \\
std & 0.087 & 0.166 & 0.082 & 0.066 & 0.120 \\
min & 0.375 & 0.157 & 0.113 & 0.447 & 0.152 \\
25\% & 0.580 & 0.504 & 0.535 & 0.541 & 0.459 \\
50\% & 0.633 & 0.609 & 0.566 & 0.579 & 0.528 \\
75\% & 0.688 & 0.688 & 0.612 & 0.632 & 0.582 \\
max & 0.834 & 0.945 & 0.787 & 0.774 & 0.765 \\
\bottomrule
\end{tabular*}
\end{table}

\FloatBarrier

\subsection{Distributions of Model Performance Scores (Transformer, Person-Specific Models)}

\begin{table}[h]
    \begin{tabular*}{\textwidth}{@{\extracolsep{\fill}}lrrrrr}
\toprule
 & auc & acc & pre & rec & f1 \\
\midrule
mean & 0.554 & 0.580 & 0.419 & 0.518 & 0.398 \\
std & 0.075 & 0.251 & 0.173 & 0.044 & 0.145 \\
min & 0.411 & 0.083 & 0.041 & 0.442 & 0.076 \\
25\% & 0.500 & 0.357 & 0.330 & 0.500 & 0.297 \\
50\% & 0.546 & 0.646 & 0.459 & 0.500 & 0.447 \\
75\% & 0.592 & 0.792 & 0.536 & 0.514 & 0.487 \\
max & 0.770 & 0.953 & 0.866 & 0.731 & 0.681 \\
\bottomrule
\end{tabular*}
\end{table}

\FloatBarrier
\newpage

\subsection{Distributions of Model Performance Scores (LSTM, Fine-Tuned Models)}

\begin{table}[h]
    \begin{tabular*}{\textwidth}{@{\extracolsep{\fill}}lrrrrr}
\toprule
 & auc & acc & pre & rec & f1 \\
\midrule
mean & 0.696 & 0.739 & 0.622 & 0.629 & 0.616 \\
std & 0.088 & 0.106 & 0.059 & 0.075 & 0.063 \\
min & 0.501 & 0.544 & 0.517 & 0.516 & 0.503 \\
25\% & 0.636 & 0.658 & 0.581 & 0.574 & 0.571 \\
50\% & 0.686 & 0.728 & 0.612 & 0.616 & 0.606 \\
75\% & 0.743 & 0.813 & 0.666 & 0.661 & 0.658 \\
max & 0.909 & 0.970 & 0.776 & 0.826 & 0.774 \\
\bottomrule
\end{tabular*}
\end{table}

\FloatBarrier

\subsection{Distributions of Model Performance Scores (Transformer, Fine-Tuned Models)}

\begin{table}[h]
    \begin{tabular*}{\textwidth}{@{\extracolsep{\fill}}lrrrrr}
\toprule
 & auc & acc & pre & rec & f1 \\
\midrule
mean & 0.682 & 0.720 & 0.608 & 0.623 & 0.605 \\
std & 0.076 & 0.104 & 0.056 & 0.064 & 0.057 \\
min & 0.520 & 0.529 & 0.499 & 0.499 & 0.495 \\
25\% & 0.635 & 0.653 & 0.574 & 0.576 & 0.566 \\
50\% & 0.679 & 0.693 & 0.601 & 0.625 & 0.604 \\
75\% & 0.731 & 0.785 & 0.643 & 0.656 & 0.641 \\
max & 0.849 & 0.966 & 0.748 & 0.779 & 0.730 \\
\bottomrule
\end{tabular*}
\end{table}

The full set of results is available for download in CSV format on this project's OSF page
(\href{https://osf.io/rkswe/}{https://osf.io/rkswe/}).

\newpage
\section{Models Trained on Same-Day Sequences}
\label{sameday}

\subsection{Performance of Global LSTM Model Across 20 Rounds of Hyperparameter Search}

\begin{table}[h]
\begin{tabular*}{\textwidth}{@{\extracolsep{\fill}}lrrrrr}
\toprule
rank & acc & pre & rec & f1 & auc \\
\midrule
1 & 0.777 & 0.707 & 0.661 & 0.675 & 0.781 \\
2 & 0.779 & 0.710 & 0.666 & 0.680 & 0.779 \\
3 & 0.777 & 0.707 & 0.666 & 0.680 & 0.777 \\
4 & 0.770 & 0.703 & 0.615 & 0.628 & 0.764 \\
5 & 0.770 & 0.702 & 0.619 & 0.632 & 0.762 \\
6 & 0.768 & 0.719 & 0.586 & 0.589 & 0.762 \\
7 & 0.770 & 0.714 & 0.600 & 0.608 & 0.761 \\
8 & 0.771 & 0.707 & 0.613 & 0.626 & 0.760 \\
9 & 0.768 & 0.697 & 0.621 & 0.634 & 0.758 \\
10 & 0.769 & 0.710 & 0.600 & 0.609 & 0.757 \\
11 & 0.769 & 0.702 & 0.611 & 0.623 & 0.757 \\
12 & 0.768 & 0.700 & 0.610 & 0.622 & 0.756 \\
13 & 0.768 & 0.696 & 0.619 & 0.633 & 0.756 \\
14 & 0.769 & 0.706 & 0.605 & 0.615 & 0.754 \\
15 & 0.768 & 0.704 & 0.605 & 0.616 & 0.755 \\
16 & 0.766 & 0.690 & 0.628 & 0.641 & 0.753 \\
17 & 0.766 & 0.693 & 0.616 & 0.629 & 0.749 \\
18 & 0.741 & 0.371 & 0.500 & 0.426 & 0.632 \\
19 & 0.741 & 0.371 & 0.500 & 0.426 & 0.500 \\
20 & 0.741 & 0.371 & 0.500 & 0.426 & 0.500 \\
\bottomrule

\end{tabular*}
\end{table}

\subsection{Performance of Global Transformer Model Across 20 Rounds of Hyperparameter Search}

\begin{table}[h]
\begin{tabular*}{\textwidth}{@{\extracolsep{\fill}}lrrrrr}
\toprule
rank & acc & pre & rec & f1 & auc \\
\midrule
1 & 0.773 & 0.701 & 0.657 & 0.670 & 0.775 \\
2 & 0.775 & 0.705 & 0.648 & 0.663 & 0.775 \\
3 & 0.775 & 0.706 & 0.644 & 0.659 & 0.774 \\
4 & 0.778 & 0.713 & 0.641 & 0.658 & 0.775 \\
5 & 0.775 & 0.717 & 0.622 & 0.636 & 0.774 \\
6 & 0.776 & 0.709 & 0.642 & 0.658 & 0.774 \\
7 & 0.776 & 0.708 & 0.644 & 0.660 & 0.774 \\
8 & 0.773 & 0.702 & 0.649 & 0.664 & 0.774 \\
9 & 0.774 & 0.703 & 0.647 & 0.662 & 0.773 \\
10 & 0.774 & 0.705 & 0.646 & 0.661 & 0.774 \\
11 & 0.776 & 0.716 & 0.624 & 0.639 & 0.773 \\
12 & 0.771 & 0.698 & 0.658 & 0.671 & 0.772 \\
13 & 0.775 & 0.711 & 0.631 & 0.647 & 0.772 \\
14 & 0.776 & 0.712 & 0.634 & 0.650 & 0.773 \\
15 & 0.775 & 0.704 & 0.652 & 0.666 & 0.772 \\
16 & 0.775 & 0.706 & 0.643 & 0.658 & 0.772 \\
17 & 0.775 & 0.714 & 0.625 & 0.640 & 0.770 \\
18 & 0.771 & 0.719 & 0.600 & 0.609 & 0.769 \\
19 & 0.769 & 0.701 & 0.617 & 0.630 & 0.764 \\
20 & 0.767 & 0.690 & 0.634 & 0.648 & 0.762 \\
\bottomrule
\end{tabular*}
\end{table}

The full set of results is available for download in CSV format on this project's OSF page
(\href{https://osf.io/rkswe/}{https://osf.io/rkswe/}).

\newpage
\section{Logistic Regression and Random Forest Models as Additional Benchmarks}
\label{LRRF}

\subsection{Performance of Best Global Logistic Regression and Random Forest Models}

\begin{table}[h]
\begin{tabular*}{\textwidth}{@{\extracolsep{\fill}}rrrrrr}
\toprule
model & acc & pre & rec & f1 & auc \\
\midrule
LR & 0.746 & 0.656 & 0.624 & 0.634 & 0.713 \\
RF & 0.761 & 0.721 & 0.562 & 0.551 & 0.732 \\

\bottomrule
\end{tabular*}
\end{table}

\FloatBarrier

\subsection{Distributions of Model Performance Scores (Logistic Regression)}

\begin{table}[h]
    \begin{tabular*}{\textwidth}{@{\extracolsep{\fill}}lrrrrr}
\toprule
 & auc & acc & pre & rec & f1 \\
\midrule
mean & 0.615 & 0.748 & 0.573 & 0.551 & 0.550 \\
std & 0.063 & 0.103 & 0.058 & 0.044 & 0.049 \\
min & 0.486 & 0.526 & 0.476 & 0.484 & 0.477 \\
25\% & 0.571 & 0.668 & 0.526 & 0.516 & 0.512 \\
50\% & 0.613 & 0.740 & 0.569 & 0.543 & 0.547 \\
75\% & 0.660 & 0.827 & 0.611 & 0.579 & 0.582 \\
max & 0.802 & 0.954 & 0.784 & 0.679 & 0.679 \\
\bottomrule
\end{tabular*}
\end{table}

\FloatBarrier

\subsection{Distributions of Model Performance Scores (Random Forest)}

\begin{table}[h]
    \begin{tabular*}{\textwidth}{@{\extracolsep{\fill}}lrrrrr}
\toprule
 & auc & acc & pre & rec & f1 \\
\midrule
count & 99.000 & 99.000 & 99.000 & 99.000 & 99.000 \\
mean & 0.647 & 0.766 & 0.590 & 0.520 & 0.474 \\
std & 0.068 & 0.124 & 0.169 & 0.034 & 0.056 \\
min & 0.515 & 0.461 & 0.316 & 0.497 & 0.386 \\
25\% & 0.604 & 0.688 & 0.452 & 0.500 & 0.439 \\
50\% & 0.640 & 0.775 & 0.594 & 0.505 & 0.462 \\
75\% & 0.677 & 0.859 & 0.681 & 0.518 & 0.493 \\
max & 0.838 & 0.979 & 0.961 & 0.654 & 0.673 \\
\bottomrule
\end{tabular*}
\end{table}

The full set of results is available for download in CSV format on this project's OSF page
(\href{https://osf.io/rkswe/}{https://osf.io/rkswe/}).

\vspace{5mm}

We used regularized logistic regression (LR) and random forest (RF) models as additional banchmarks. To generate the feature set, we chose a moving window approach, mirroring the approach chosen to train the neural network models as closely as possible. We used a moving time window of 20 app sessions per data point in alignment with the sequence length of 20 used for the neural network models. Each preceding app session was mapped onto one of the input features. For example, t-1 was always the first feature while t-20 was always the last. This is a standard approach when applying classic ML approaches like LR or RF to sequence data. We performed hyperparameter search and selected the best model on the validation set before estimating model performance on the test set. All splits were held constant with respect to those used for the neural network models to ensure a fair comparison.

Results show that both the LR and the RF models perform better than chance but lower than the neural network models. The performance of the LR on the main prediction task was $AUC_{lr}$ = 0.713 and the performance of the RF was $AUC_{rf}$ = 0.732. These results are in the range of neural network models being trained on only one to two preceding app sessions but substantially lower than those of full neural network models ($AUC_{trans}$=0.773, $AUC_{lstm}$=0.782), hence lending credence to the interpretation that the neural network models pick up on sequential relationships above and beyond the capacity of conventional classification models in the present predictive task. Detailed results can be found in the table above. 

\FloatBarrier
\newpage

\section{Fine-Tuned Models (Top Layers Only)}
\label{habits_finetuned}

\subsection{Distributions of Model Performance Scores (LSTM)}
\begin{table}[h]
    \begin{tabular*}{\textwidth}{@{\extracolsep{\fill}}lrrrrr}
\toprule
 & auc & acc & pre & rec & f1 \\
\midrule
mean & 0.678 & 0.771 & 0.532 & 0.554 & 0.521 \\
std & 0.090 & 0.118 & 0.147 & 0.081 & 0.109 \\
min & 0.394 & 0.251 & 0.285 & 0.483 & 0.251 \\
25\% & 0.626 & 0.689 & 0.410 & 0.500 & 0.447 \\
50\% & 0.678 & 0.782 & 0.496 & 0.500 & 0.481 \\
75\% & 0.723 & 0.856 & 0.638 & 0.601 & 0.599 \\
max & 0.902 & 0.978 & 0.906 & 0.801 & 0.805 \\
\bottomrule
\end{tabular*}
\end{table}

\FloatBarrier
\vspace{0.5cm}

\subsection{Distributions of Model Performance Scores (Transformer)}

\begin{table}[h]
    \begin{tabular*}{\textwidth}{@{\extracolsep{\fill}}lrrrrr}
\toprule
 & auc & acc & pre & rec & f1 \\
\midrule
mean & 0.654 & 0.766 & 0.585 & 0.554 & 0.538 \\
std & 0.083 & 0.111 & 0.116 & 0.065 & 0.085 \\
min & 0.384 & 0.517 & 0.369 & 0.406 & 0.405 \\
25\% & 0.599 & 0.681 & 0.496 & 0.500 & 0.475 \\
50\% & 0.655 & 0.766 & 0.589 & 0.532 & 0.524 \\
75\% & 0.703 & 0.846 & 0.650 & 0.592 & 0.590 \\
max & 0.819 & 0.981 & 0.915 & 0.767 & 0.776 \\
\bottomrule
\end{tabular*}
\end{table}

The full set of results is available for download in CSV format on this project's OSF page
(\href{https://osf.io/rkswe/}{https://osf.io/rkswe/}).

\FloatBarrier
\newpage

\newpage
\section{Examples of App N-Grams with Empirical Transition Probabilities}
\label{ngrams}
\FloatBarrier

\begin{table}[h]
    \begin{tabular*}{\textwidth}{@{\extracolsep{\fill}}lrr}
\toprule
n-grams & transition  & n-gram  \\
 &  probability & frequency \\

\midrule
Snapchat, Snapchat, Snapchat & 0.81 & 2691 \\
Google Chrome, Google Chrome, Google Chrome & 0.13 & 1430 \\
Samsung Messaging, Samsung Messaging, Samsung Messaging & 0.10 & 1425 \\
Google Search, Google Chrome, Google Search & 0.16 & 1155 \\
Facebook, Facebook, Facebook & 0.51 & 924 \\
Facebook, Messenger, Facebook & 0.17 & 916 \\
Messenger, Messenger, Messenger & 0.18 & 905 \\
Reddit, Google Chrome, Reddit & 0.19 & 891 \\
Google Chrome, Google Search, Google Chrome & 0.10 & 879 \\
Messenger, Facebook, Messenger & 0.57 & 786 \\
Instagram, Instagram, Instagram & 0.67 & 721 \\
Google Chrome, Reddit, Google Chrome & 0.77 & 717 \\
Reddit, Reddit, Reddit & 0.73 & 642 \\
Google Mail, Google Chrome, Google Mail & 0.20 & 607 \\
Google Messaging, Google Messaging, Google Messaging & 0.17 & 581 \\
Google Chrome, Google Mail, Google Chrome & 0.12 & 540 \\
Discord, Discord, Discord & 0.67 & 492 \\
Samsung Messaging, Google Chrome, Samsung Messaging & 0.27 & 475 \\
Discord, Google Chrome, Discord & 0.19 & 406 \\
Messenger, Messenger, Facebook & 0.45 & 385 \\
\bottomrule
\end{tabular*}

\vspace{1mm}
Most frequent 3-grams and their social media transition probabilities across all participants. 

\end{table}

\FloatBarrier

\begin{table}[h]
    \begin{tabular*}{\textwidth}{@{\extracolsep{\fill}}lrr}
\toprule
n-grams & transition  & n-gram  \\
 &  probability & frequency \\
 \midrule
Samsung Messaging, Instagram, Samsung Messaging & 0.41 & 203 \\
Samsung Messaging, Google Play Store, Samsung Messaging & 0.16 & 158 \\
Instagram, Samsung Messaging, Samsung Messaging & 0.58 & 112 \\
Google Play Store, Samsung Messaging, Google Play Store & 0.03 & 108 \\
Instagram, Samsung Messaging, Instagram & 0.02 & 97 \\
Samsung Messaging, Samsung Messaging, Instagram & 0.04 & 80 \\
Samsung Messaging, Google Chrome, Samsung Messaging & 0.42 & 71 \\
Samsung Messaging, Twitter, Samsung Messaging & 0.61 & 61 \\
Samsung Messaging, Samsung Messaging, Samsung Messaging & 0.18 & 57 \\
Samsung Messaging, AccuWeather, Samsung Messaging & 0.25 & 53 \\
Samsung Messaging, Weather Channel, Samsung Messaging & 0.22 & 49 \\
Twitter, Samsung Messaging, Twitter & 0.15 & 40 \\
Samsung Messaging, Robinhood, Samsung Messaging & 0.29 & 35 \\
Samsung Messaging, Reddit, Samsung Messaging & 0.42 & 26 \\
AccuWeather, Samsung Messaging, AccuWeather & 0.05 & 20 \\
Samsung Messaging, Samsung Messaging, Google Chrome & 0.10 & 20 \\
Samsung Messaging, Samsung Messaging, Google Play Store & 0.10 & 20 \\
Twitter, Google Chrome, Twitter & 0.26 & 19 \\
Robinhood, Samsung Messaging, Robinhood & 0.16 & 19 \\
Samsung Messaging, Samsung Messaging, AccuWeather & 0.00 & 18 \\
\bottomrule
\end{tabular*}

\vspace{1mm}
Most frequent 3-grams and their social media transition probabilities for user b\_118. 

\end{table}

\FloatBarrier

\begin{table}[h]
    \begin{tabular*}{\textwidth}{@{\extracolsep{\fill}}lrr}
\toprule
n-grams & transition  & n-gram  \\
 &  probability & frequency \\
 \midrule
Google Search, Google Chrome, Google Search & 0.17 & 36 \\
Google Chrome, Google Search, Google Chrome & 0.18 & 34 \\
Reddit, Google Messaging, Google Chrome & 0.44 & 16 \\
Google Messaging, Google Chrome, Reddit & 0.21 & 14 \\
Facebook, Google Messaging, Google Chrome & 0.64 & 11 \\
Facebook, Google Docs, Facebook & 0.20 & 10 \\
Reddit, Google Search, Google Chrome & 0.11 & 9 \\
Google Mail, Google Chrome, Google Mail & 0.62 & 8 \\
Google Docs, Google Mail, Google Docs & 0.00 & 7 \\
Reddit, Google Chat, Reddit & 0.14 & 7 \\
Facebook, Google Search, Google Chrome & 0.14 & 7 \\
Reddit, Google Chrome, Reddit & 0.14 & 7 \\
Google Mail, Google Docs, Google Mail & 0.00 & 7 \\
Camera, Photos, Google Messaging & 0.43 & 7 \\
Camera, Photos, Google Chat & 0.29 & 7 \\
Camera, Photos, Camera & 0.00 & 7 \\
Google Messaging, Google Chrome, Facebook & 0.29 & 7 \\
Reddit, YouTube, Reddit & 0.33 & 6 \\
Google Chat, Camera, Google Chat & 0.00 & 6 \\
Reddit, Project Makeover, Project Makeover & 0.67 & 6 \\
\bottomrule
\end{tabular*}

\vspace{1mm}
Most frequent 3-grams and their social media transition probabilities for user b\_124. 
\end{table}

\FloatBarrier

\begin{table}[h]
    \begin{tabular*}{\textwidth}{@{\extracolsep{\fill}}lrr}
\toprule
n-grams & transition  & n-gram  \\
 &  probability & frequency \\
 \midrule
Samsung Messaging, Google Chrome, Samsung Messaging & 0.36 & 25 \\
Google Mail, Facebook, Facebook & 0.26 & 23 \\
Twitter, Google Chrome, Twitter & 0.06 & 18 \\
Samsung Messaging, Facebook, Samsung Messaging & 0.44 & 16 \\
Facebook, Samsung Messaging, Facebook & 0.19 & 16 \\
Google Mail, Google Mail, Facebook & 0.62 & 16 \\
Samsung Kids Mode, Galaxy Store,  & 0.00 & 12 \\
Facebook, Messenger, Facebook & 0.08 & 12 \\
Facebook, Google Mail, Facebook & 0.55 & 11 \\
Samsung Gallery, Samsung Photo Editor, Samsung Gallery & 0.00 & 11 \\
Facebook, Facebook, Facebook & 0.36 & 11 \\
Google Chrome, Google Chrome, Google Chrome & 0.00 & 10 \\
Facebook, Google Search, Facebook & 0.10 & 10 \\
Twitter, Facebook, Twitter & 0.40 & 10 \\
Samsung Messaging, Amazon Kindle, Samsung Messaging & 0.20 & 10 \\
Google Mail, Facebook, Google Mail & 0.60 & 10 \\
Samsung Kids Mode, Lego Duple World, Samsung Kids Mode & 0.00 & 10 \\
Google Mail, Facebook, Samsung Messaging & 0.30 & 10 \\
Google Mail, Facebook, Twitter & 0.22 & 9 \\
Facebook, Samsung Messaging, Google Chrome & 0.00 & 9 \\
\bottomrule
\end{tabular*}

\vspace{1mm}
Most frequent 3-grams and their social media transition probabilities for user b\_125. 
\end{table}

\FloatBarrier

Transition probabilities were calculated as the relative frequency of social media use immediately following each 3-gram sequence across all instances of that sequence. A high value, therefore, indicates that a sequence was associated with subsequent social media use, whereas a low value indicates that a sequence was rarely followed by social media use. 

\FloatBarrier